\documentclass{elsart}

\usepackage{epsfig}

\usepackage{amssymb}

\usepackage{url}

\begin{document}

\begin{frontmatter}



{\hfill BNL-75079-2005-JA}

\title{Comparison of inclusive particle production   in 14.6 GeV/c proton-nucleus collisions with simulation}


\author[BNL]{D.~E.~Jaffe},
\author[SUNY]{K.~H.~Lo},
\author[ASU]{J.~R.~Comfort}, 
and
\author[BNL]{M.~Sivertz}

\address[ASU]{Arizona State University,  Dept. of Physics and Astronomy,  Tempe, AZ}
\address[BNL]{Brookhaven National Laboratory, Upton, NY }
\address[SUNY]{Stony Brook University,  Dept. of Physics and Astronomy, Stony Brook, NY}

\date{  \today}

\begin{abstract}
Inclusive charged pion, kaon, proton, and deuteron production
in 14.6 GeV/c proton-nucleus collisions measured by BNL experiment E802 
is compared with results from the GEANT3, GEANT4, and FLUKA 
simulation packages. The FLUKA package is found to have the best overall agreement.
\end{abstract}

\end{frontmatter}

\section{Introduction}
\label{sec:intro}

  The simulation of particle production in proton-nucleus collisions is
important for a number of ongoing and future high energy physics experiments.
Interpretation of atmospheric neutrino data requires knowledge of hadronic
interactions on light nuclei for lab energies from 1 to $10^5$ GeV~\cite{ref:stanev}.
The design of secondary beams for the study of
neutrino interactions~\cite{ref:neutrino} 
or rare kaon decay~\cite{ref:kopio} 
relies on the accurate
simulation of proton-nucleus collisions at energies in the range 10-100
GeV.
In addition, validation of simulations in accessible
energy regions is important for the interpretation of LHC data~\cite{ref:LHC}.

  In this paper we compare the data of BNL experiment E802~\cite{ref:e802} with
simulated results 
of the GEANT3~\cite{ref:GEANT3}, GEANT4~\cite{ref:GEANT4}, and
FLUKA~\cite{ref:FLUKA} packages. Experiment E802 measured $\pi^\pm$, $K^\pm$, proton, 
and deuteron production 
in the angular range $5^\circ$ to $58^\circ$ 
in collisions of 14.6 ${\rm GeV}/c$ protons with Be, Al, Cu, and Au targets.
The E802 magnetic spectrometer
had a geometrical solid angle acceptance of 25 msr and was rotated to take data at
five overlapping angular settings. Particle identification was accomplished with time-of-flight 
and a gas Cherenkov detector. The measured spectra were presented as invariant cross sections
$\frac{d^2\sigma}{2\pi m_{\rm t} dm_{\rm t} dy}$ as a function of transverse kinetic energy
$(m_{\rm t} - m_0)c^2 = \sqrt{ (m_0c^2)^2 + (p_\perp c)^2} - m_0c^2$ in bins of rapidity where
$m_0$ is the particle mass. The overall uncertainty in the cross section normalization is
estimated to be $\pm(10-15)\%$.

  The E802 results have previously been compared to simulation. The JAM1.0 hadronic 
cascade model~\cite{ref:JAM} showed good agreement with the measured 
proton, $\pi^\pm$, and $K^\pm$ spectra for all four targets as a function
of $m_{\rm t} - m_0$ and rapidity. The $p$-Be data as a function of rapidity has been
compared with FLUKA~\cite{Battistoni:2002ew} in the calculation of atmospheric neutrino flux.
The agreement is reasonable with the largest deviation being a factor $\sim\!1.2$ ($\sim\!2$) 
for the pion (kaon) spectra. Several simulation models were compared with 
the $p$-Be data as a function of  $m_{\rm t} - m_0$ and rapidity in the
framework of the CORSIKA program~\cite{ref:CORSIKA}. Pion production in 
FLUKA 2002 and UrQMD 1.3~\cite{ref:UrQMD} 
had the same slope as function of  $m_{\rm t} - m_0$ as the data over the whole rapidity
range, while the GHEISHA 2002~\cite{ref:GHEISHA}, QGSJET 01~\cite{ref:QGSJET}, and neXus 3~\cite{ref:neXus} models were unable to reproduce the slope as a function of  $m_{\rm t} - m_0$ 
over the full kinematic range of the data.

\section{Simulation packages}
\label{sec:sim}
 In this paper for the GEANT3 simulation we used the 
hadronic simulation package GCALOR version 1.05/03~\cite{ref:GCALOR}
with GEANT version 3.21, 
for the GEANT4 simulation we used GEANT version 4.7.1 and simulation
packages (``physics lists'') 
QGSP, QGSC,  QGSP\_BIC, and QGSC\_LEAD\_HP~\cite{ref:g4had}, and
we used version  2005.6 of FLUKA. 
The GEANT4 physics list QGSP employs a ``quark gluon string model... 
and a pre-equilibrium decay model with an extensive 
evaporation phase to model the subsequent nuclear 
fragmentation'' and 
is recommended~\cite{ref:g4had} for 
medium energy (15-50 GeV) protons on light targets. 
QGSC is similar to QGSP for the initial reaction and
`` ...uses chiral invariant phase-space decay ...
to model the behavior of the system's fragmentation.''
QGSP\_BIC is similar to QGSP but uses the binary cascade 
for nucleon interactions below 3 GeV. QGSC and QGSP\_BIC are
recommended physics lists for high energy applications.
The physics list  QGSC\_LEAD\_HP is recommended for the calculation
of LHC detector neutron fluxes.

 We simulated 14.6 GeV/$c$ proton interactions on  
Be, Al, Cu, and Au targets of thickness 
1478 mg/cm$^2$, 1620  mg/cm$^2$, 1434  mg/cm$^2$, and 1000  mg/cm$^2$,
respectively. The kinematics of 
charged pions, kaons, protons, and deuterons at a radius of
25 cm from the interaction point were recorded. The lifetime
of the charged mesons was artificially set to be infinite to
avoid performing a decay-in-flight correction to the measured yields.
For each target and model we generated 50 million incident protons.
No importance weighting or event biasing was used and the 
uncertainty in the evaluated cross sections is based on the
statistics of the generated events only.

\section{Comparison with E802 data}
\label{sec:comparison}

\begin{figure}\begin{center}
\epsfig{file=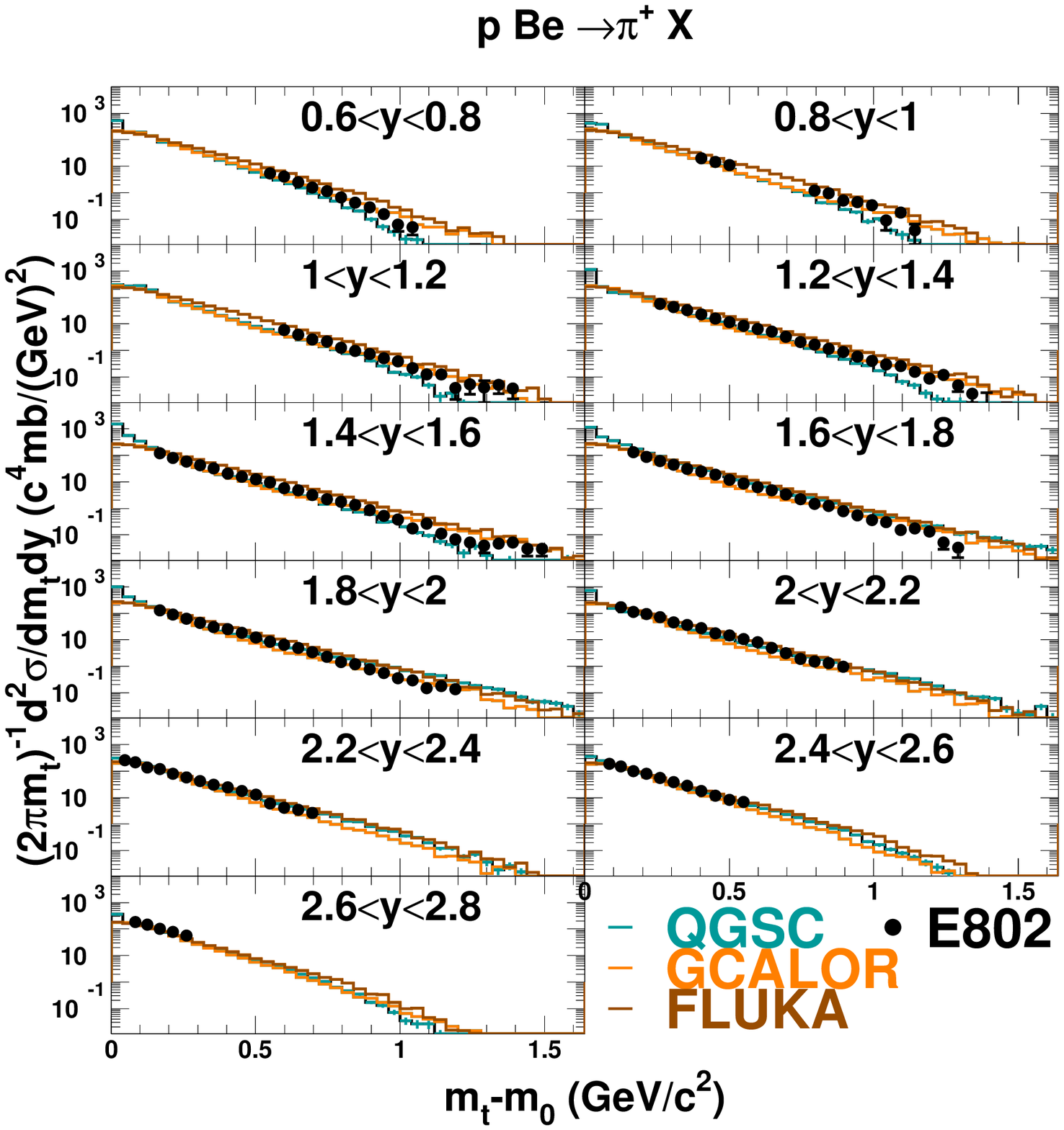,width=.495\linewidth}
\epsfig{file=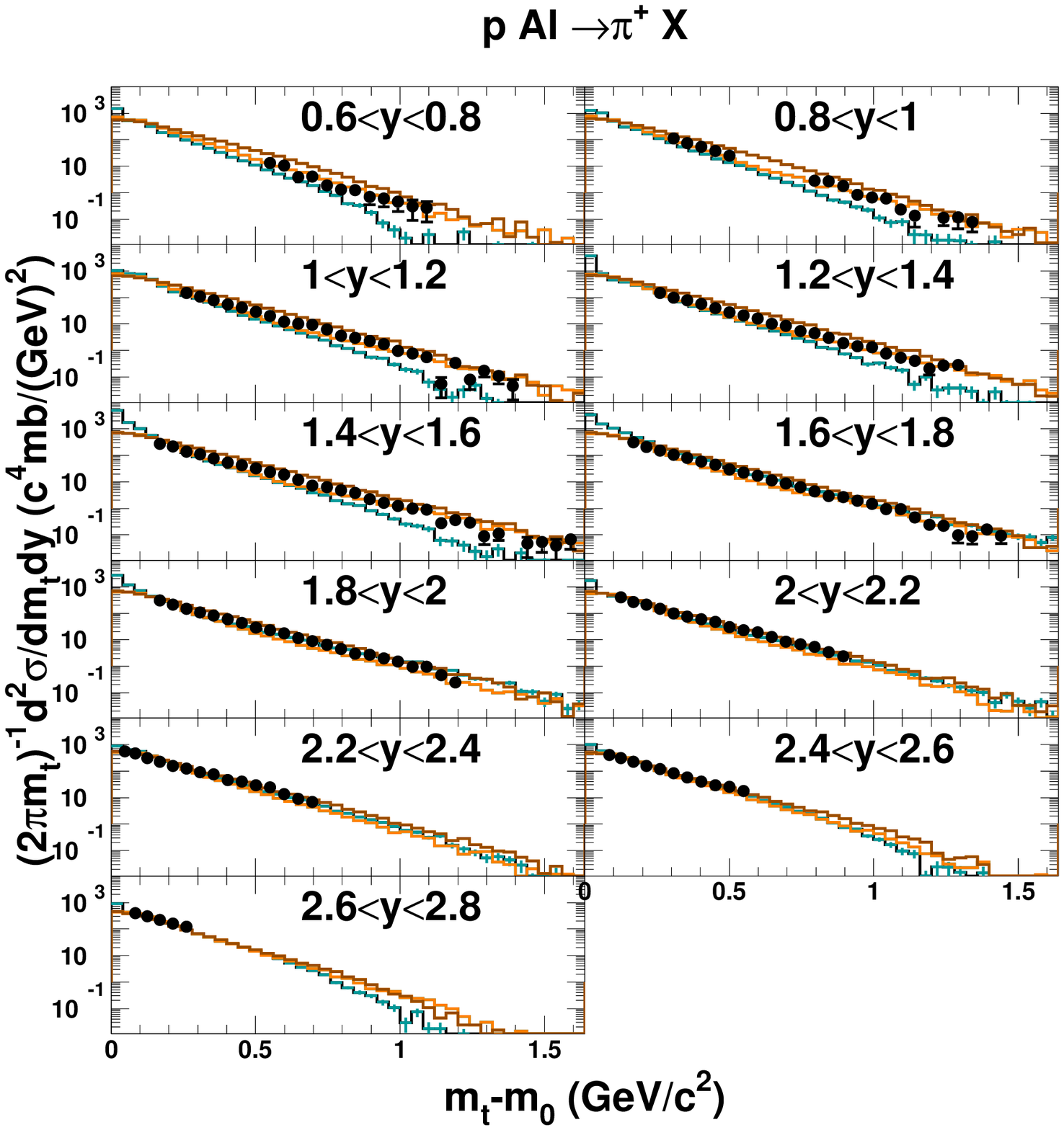,width=.495\linewidth}
\epsfig{file=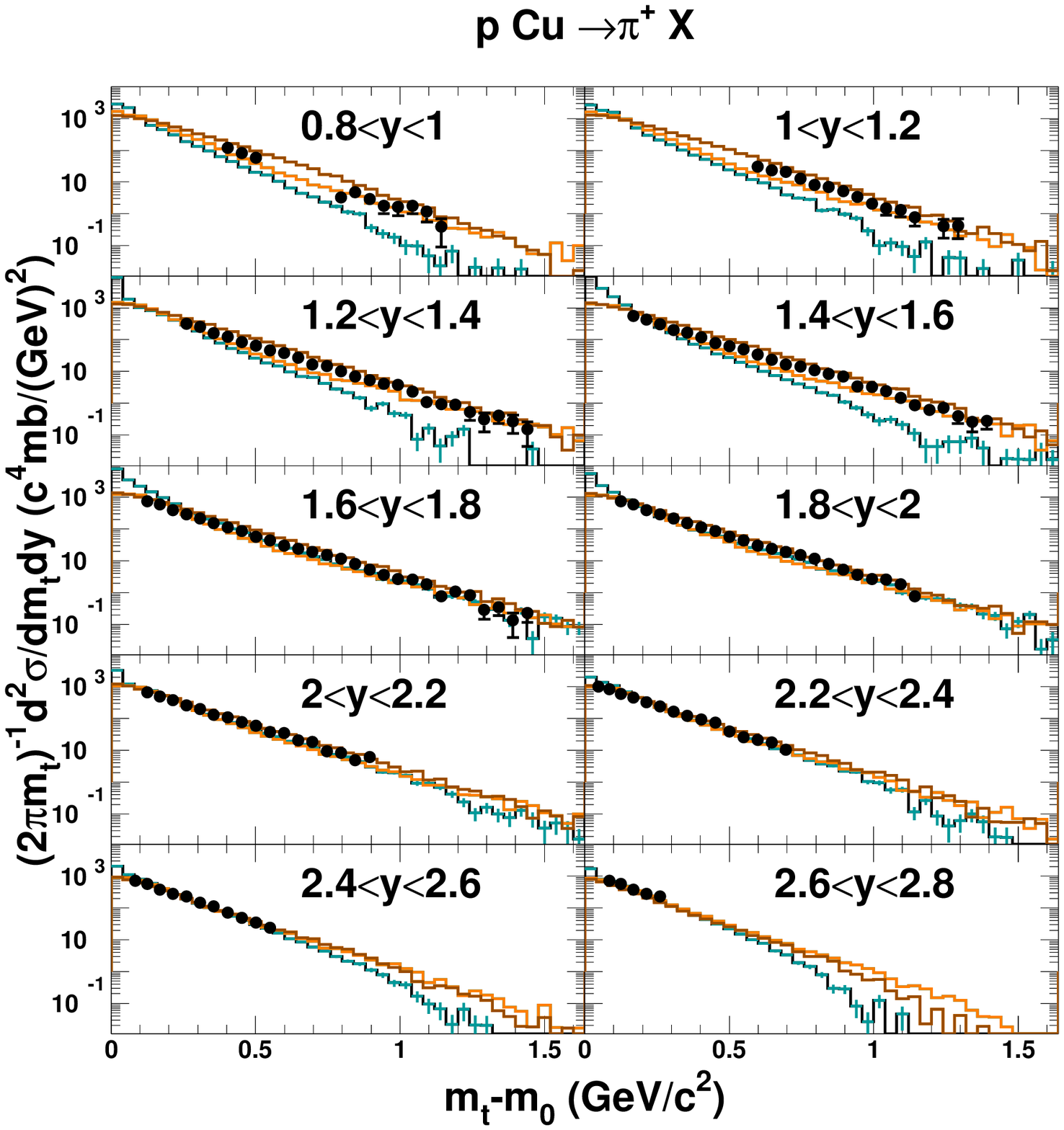,width=.495\linewidth}
\epsfig{file=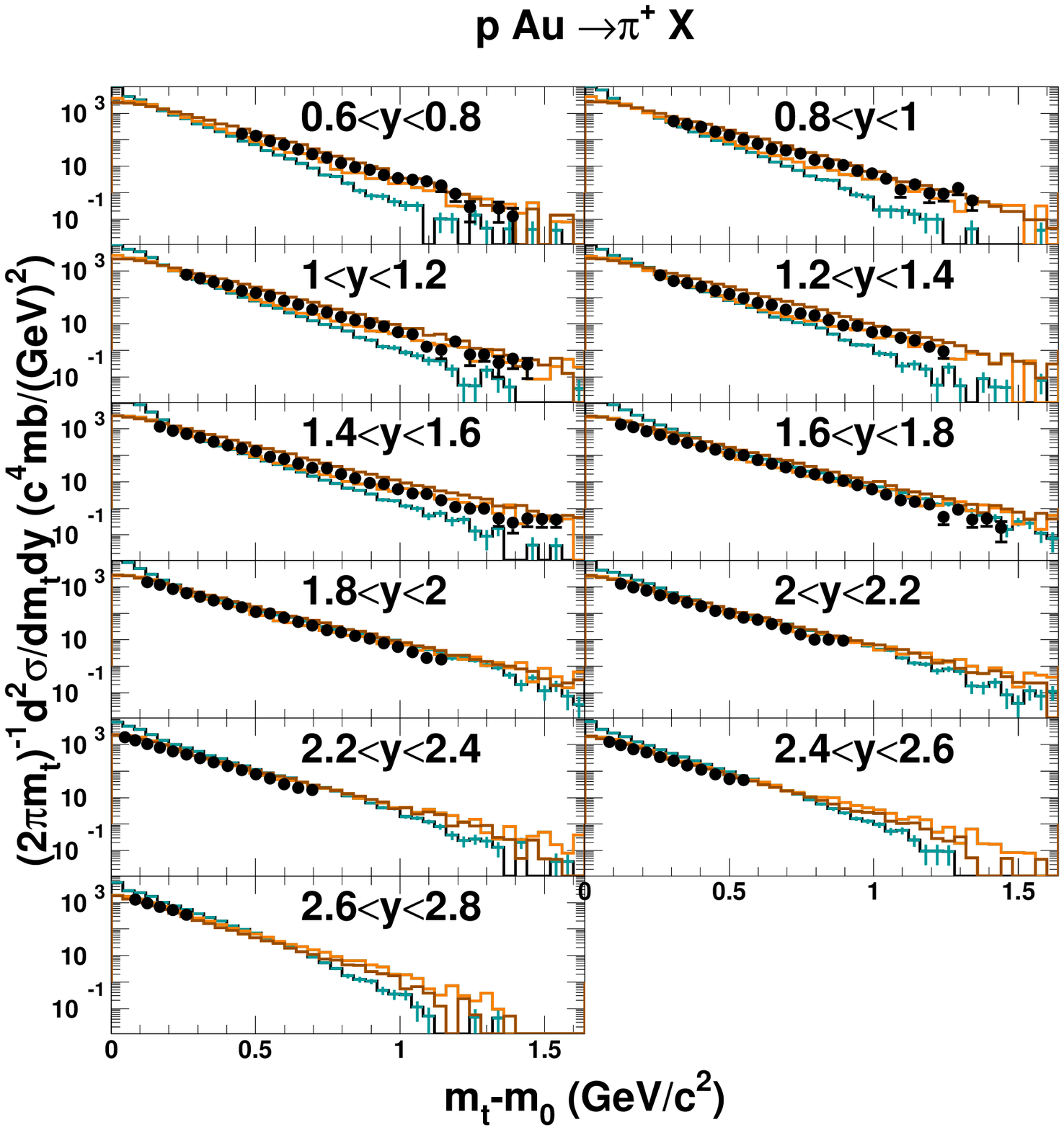,width=.495\linewidth}
\caption{\label{overlay.pi+}
The invariant cross section 
$\frac{d^2\sigma}{2\pi m_{\rm t} dm_{\rm t} dy}$ as a function of transverse kinetic energy $m_{\rm t} - m_0$ 
in 0.2 bins of rapidity compared to the simulation results for the 
$\pi^+$ data 
for $p$-Be, $p$-Al, $p$-Cu, and $p$-Au collisions. The statistical uncertainties for the different models is similar and is only shown for QGSC.}
\end{center}\end{figure}

\begin{figure}\begin{center}
\epsfig{file=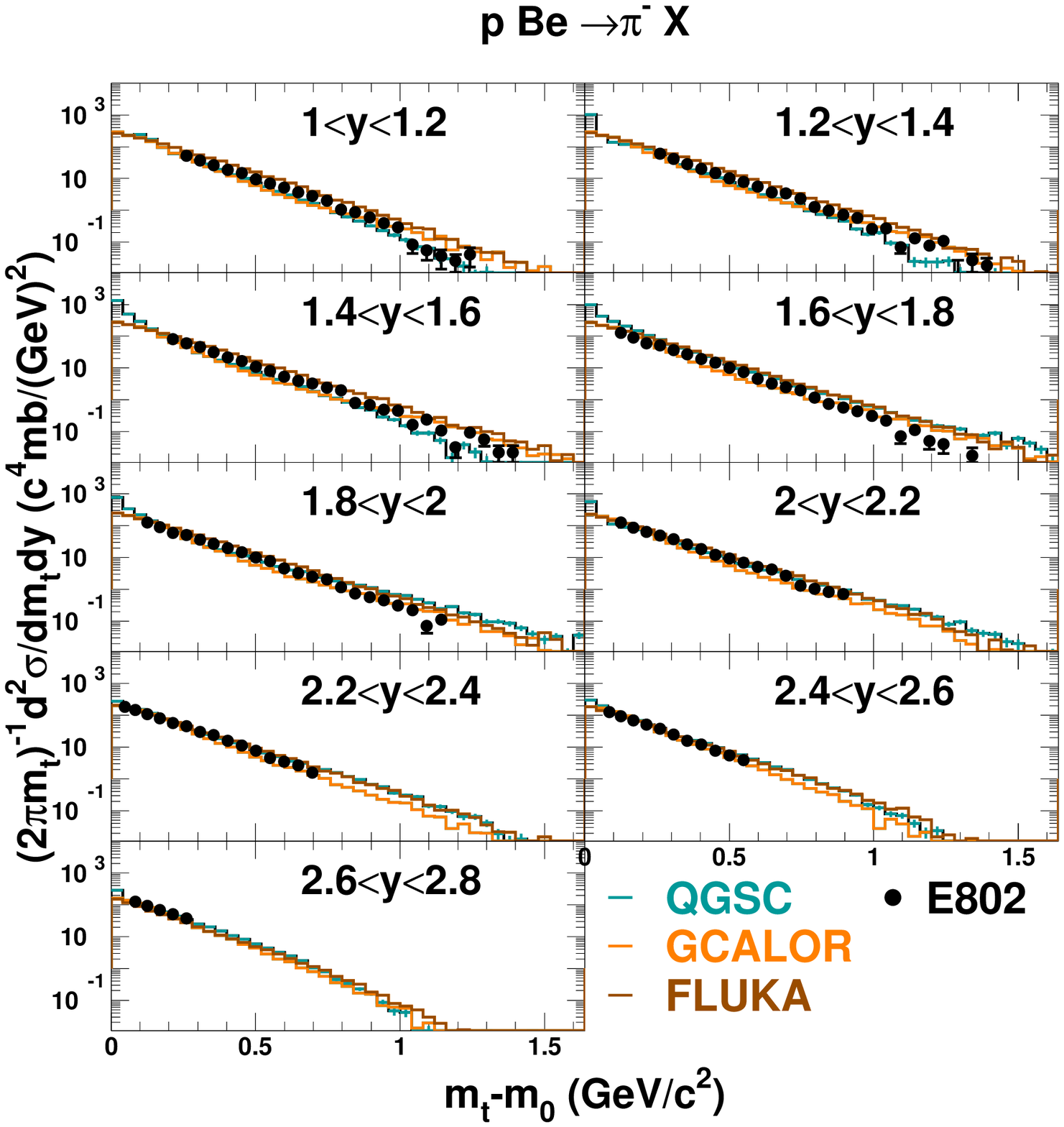,width=.495\linewidth}
\epsfig{file=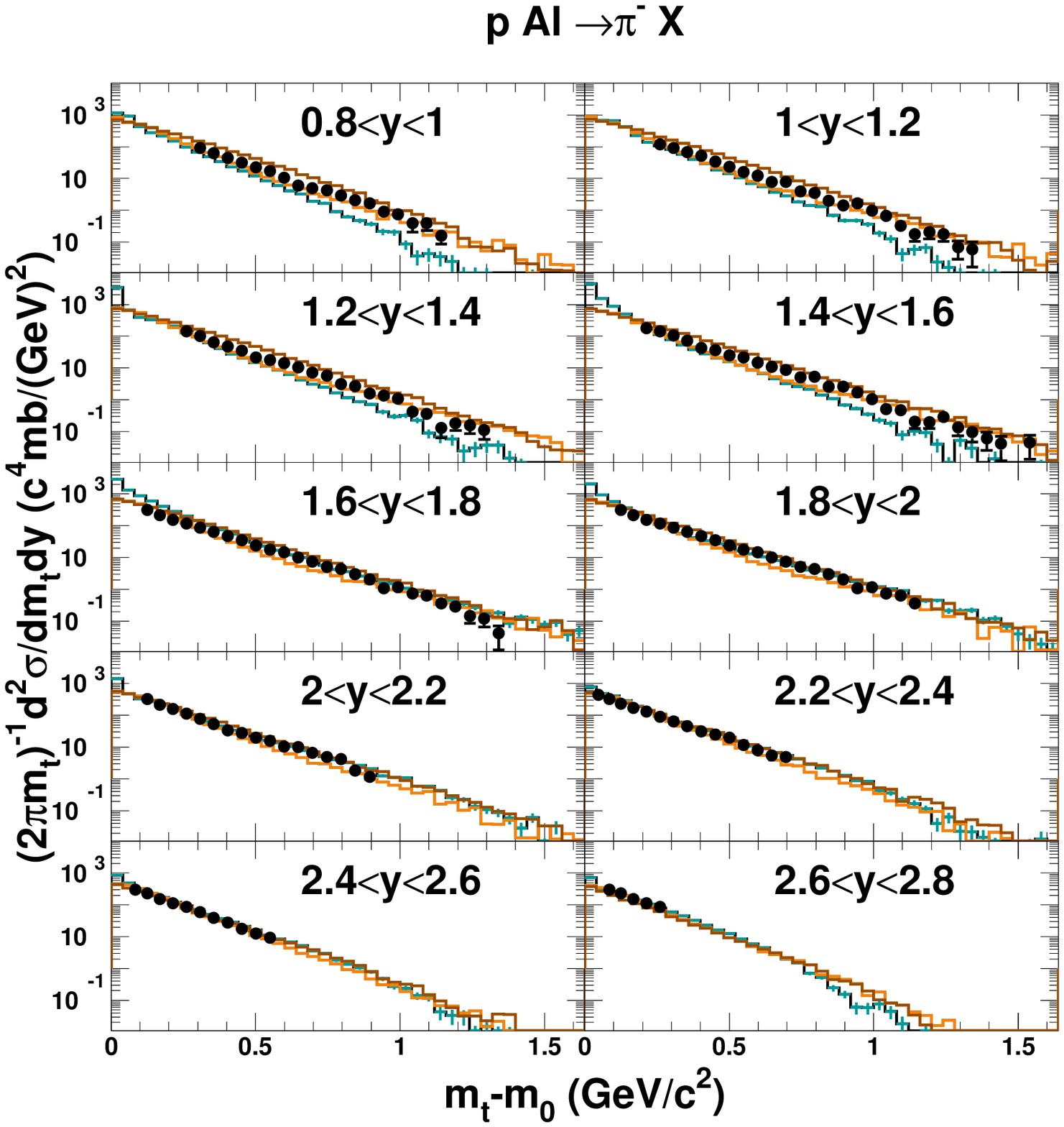,width=.495\linewidth}
\epsfig{file=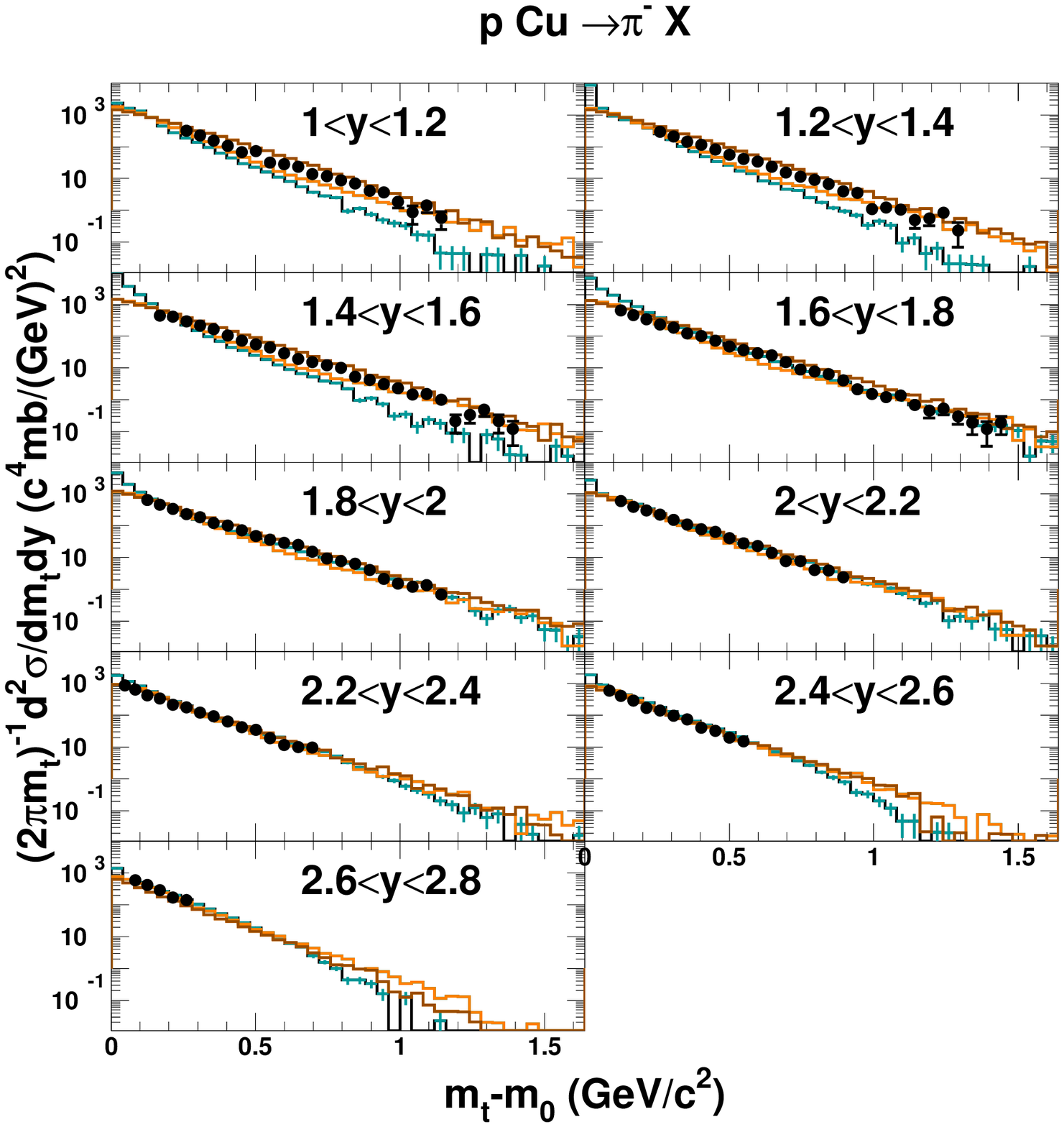,width=.495\linewidth}
\epsfig{file=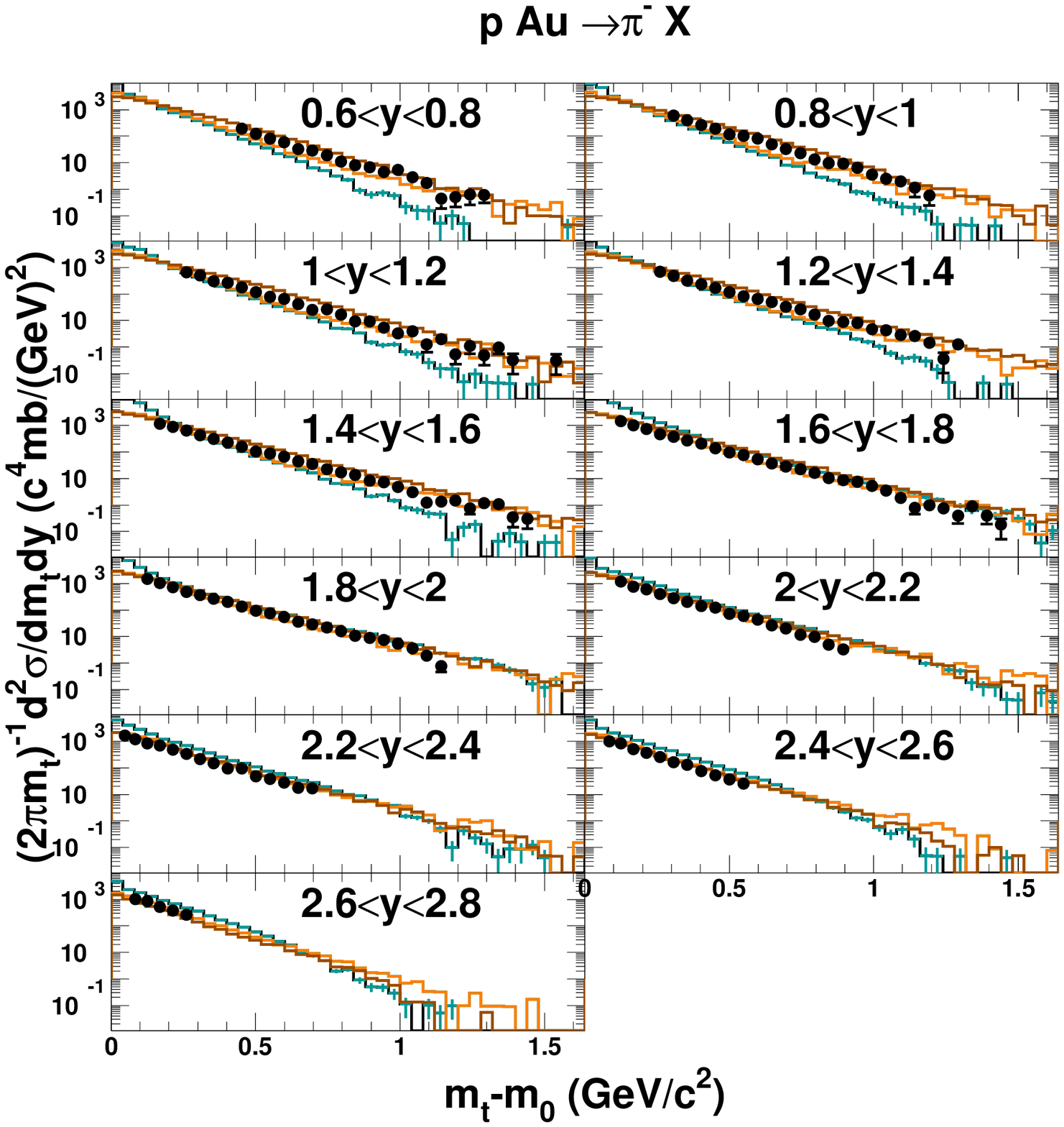,width=.495\linewidth}
\caption{\label{overlay.pi-}
The invariant cross section 
$\frac{d^2\sigma}{2\pi m_{\rm t} dm_{\rm t} dy}$ as a function of transverse kinetic energy $m_{\rm t} - m_0$ 
in 0.2 bins of rapidity compared to the simulation results for the 
$\pi^-$ data
for $p$-Be, $p$-Al, $p$-Cu, and $p$-Au collisions. The statistical uncertainties for the different models is similar and is only shown for QGSC.}
\end{center}\end{figure}

\begin{figure}\begin{center}
\epsfig{file=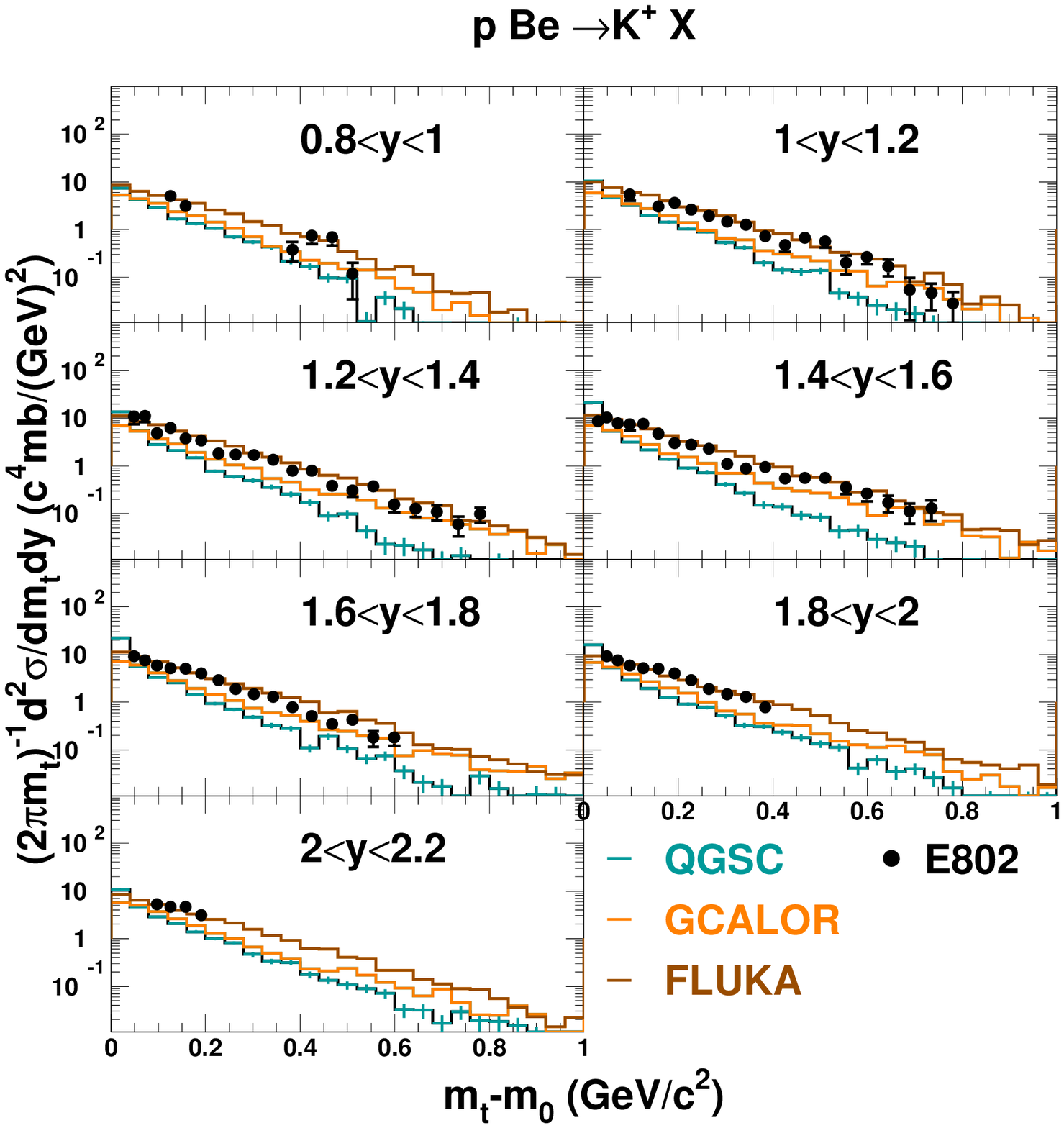,width=.495\linewidth}
\epsfig{file=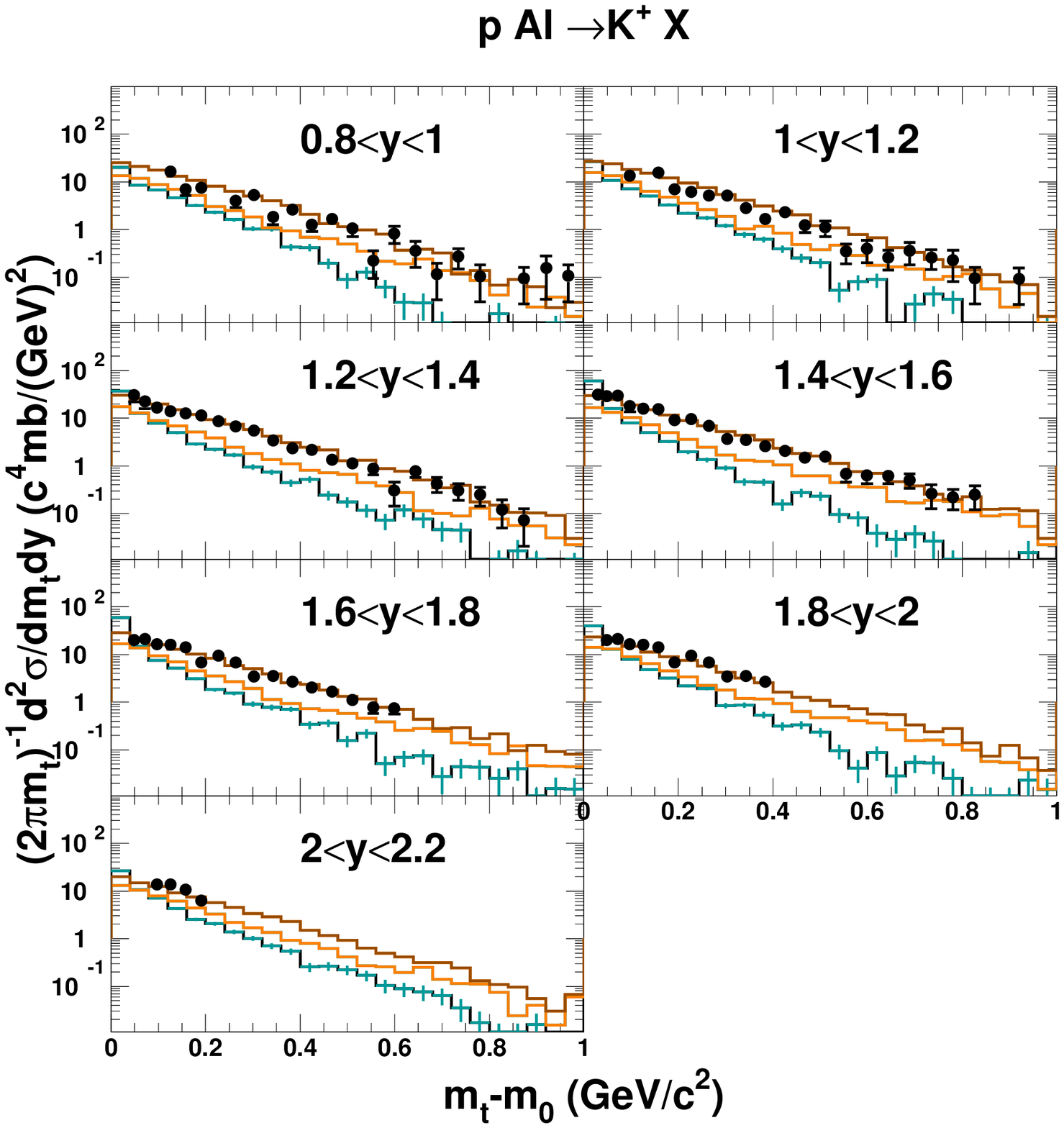,width=.495\linewidth}
\epsfig{file=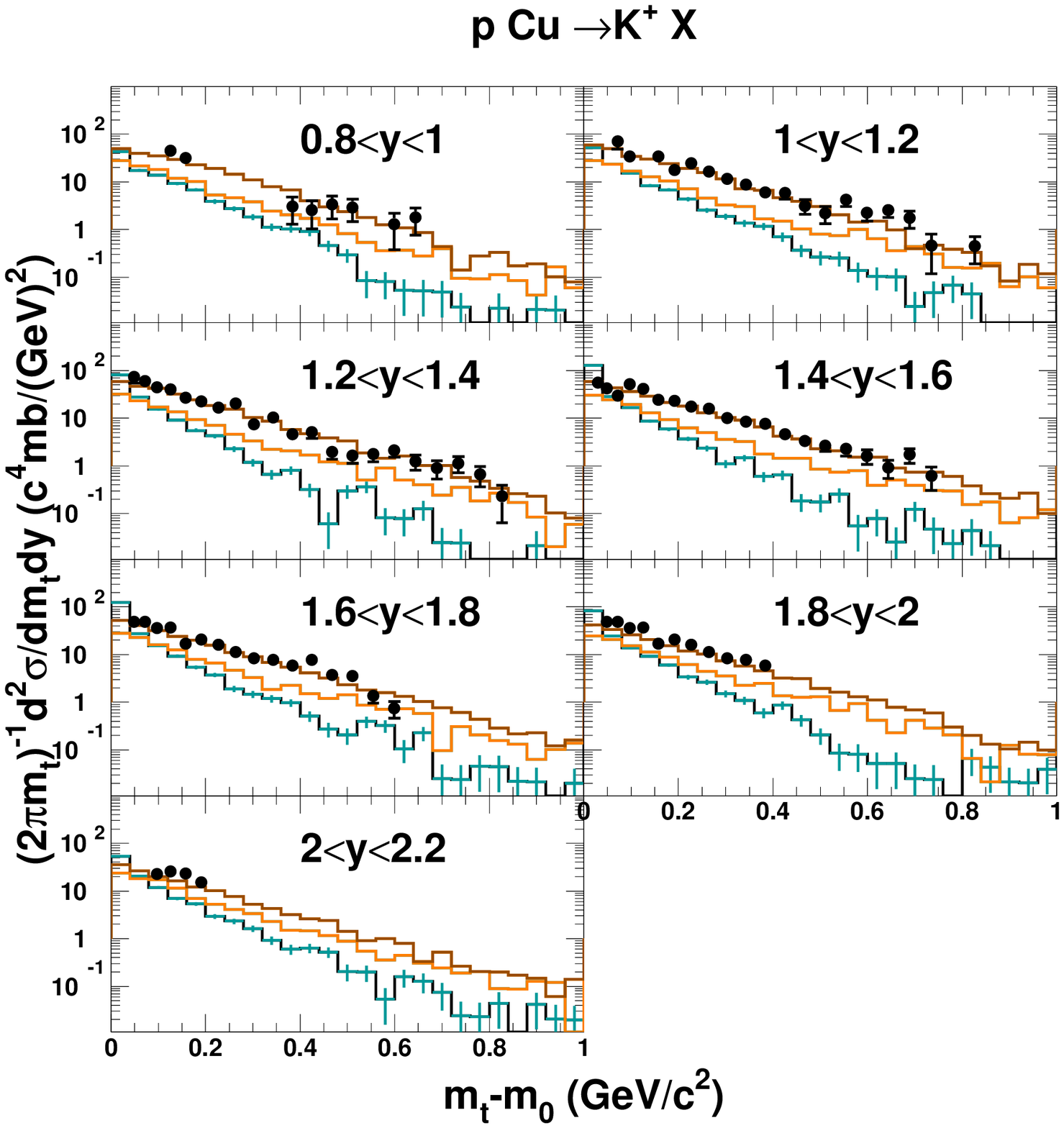,width=.495\linewidth}
\epsfig{file=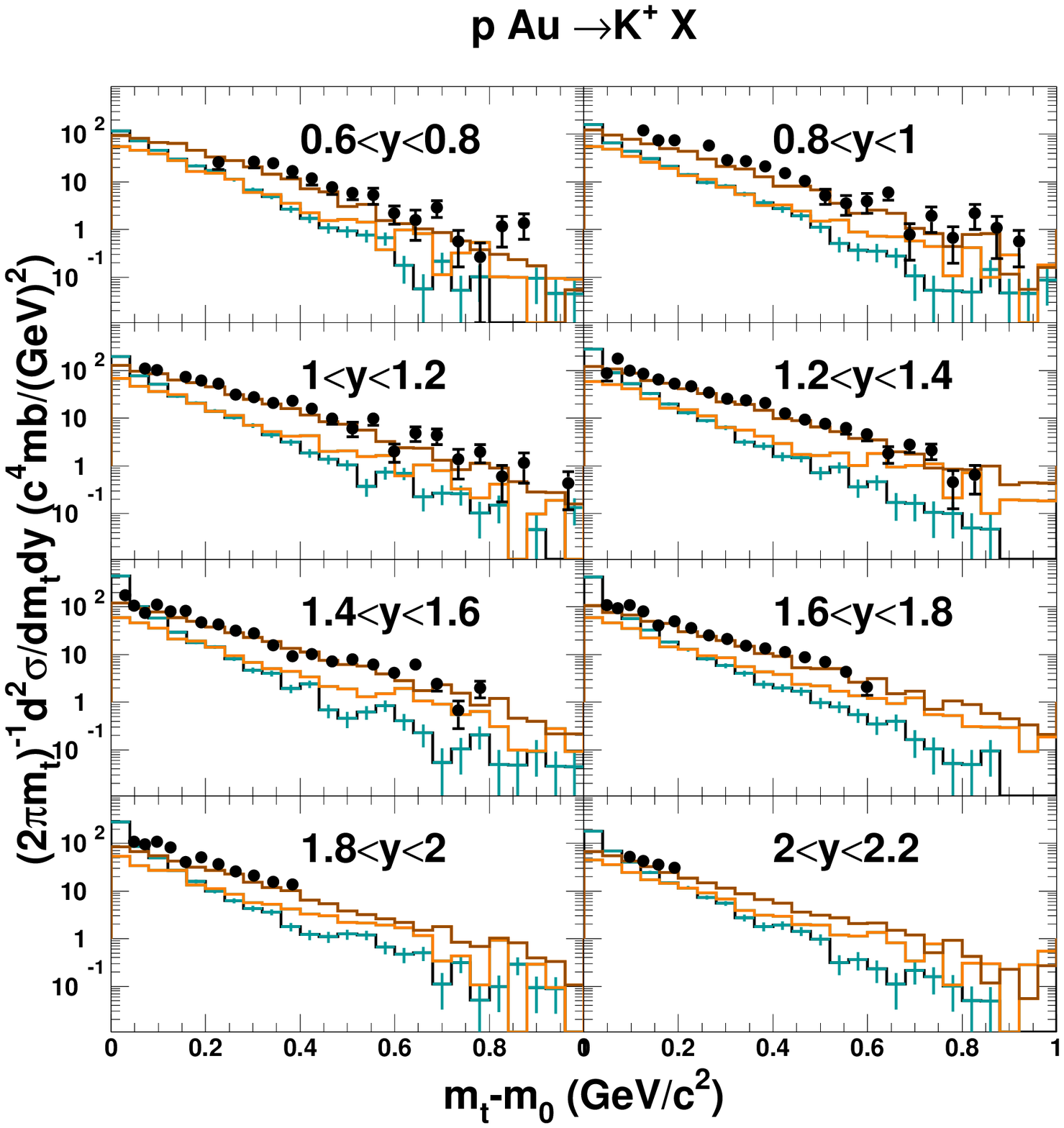,width=.495\linewidth}
\caption{\label{overlay.K+}
The invariant cross section 
$\frac{d^2\sigma}{2\pi m_{\rm t} dm_{\rm t} dy}$ as a function of transverse kinetic energy $m_{\rm t} - m_0$ 
in 0.2 bins of rapidity compared to the simulation results for the 
$K^+$ data
for $p$-Be, $p$-Al, $p$-Cu, and $p$-Au collisions. The statistical uncertainties for the different models is similar and is only shown for QGSC.}
\end{center}\end{figure}

\begin{figure}\begin{center}
\epsfig{file=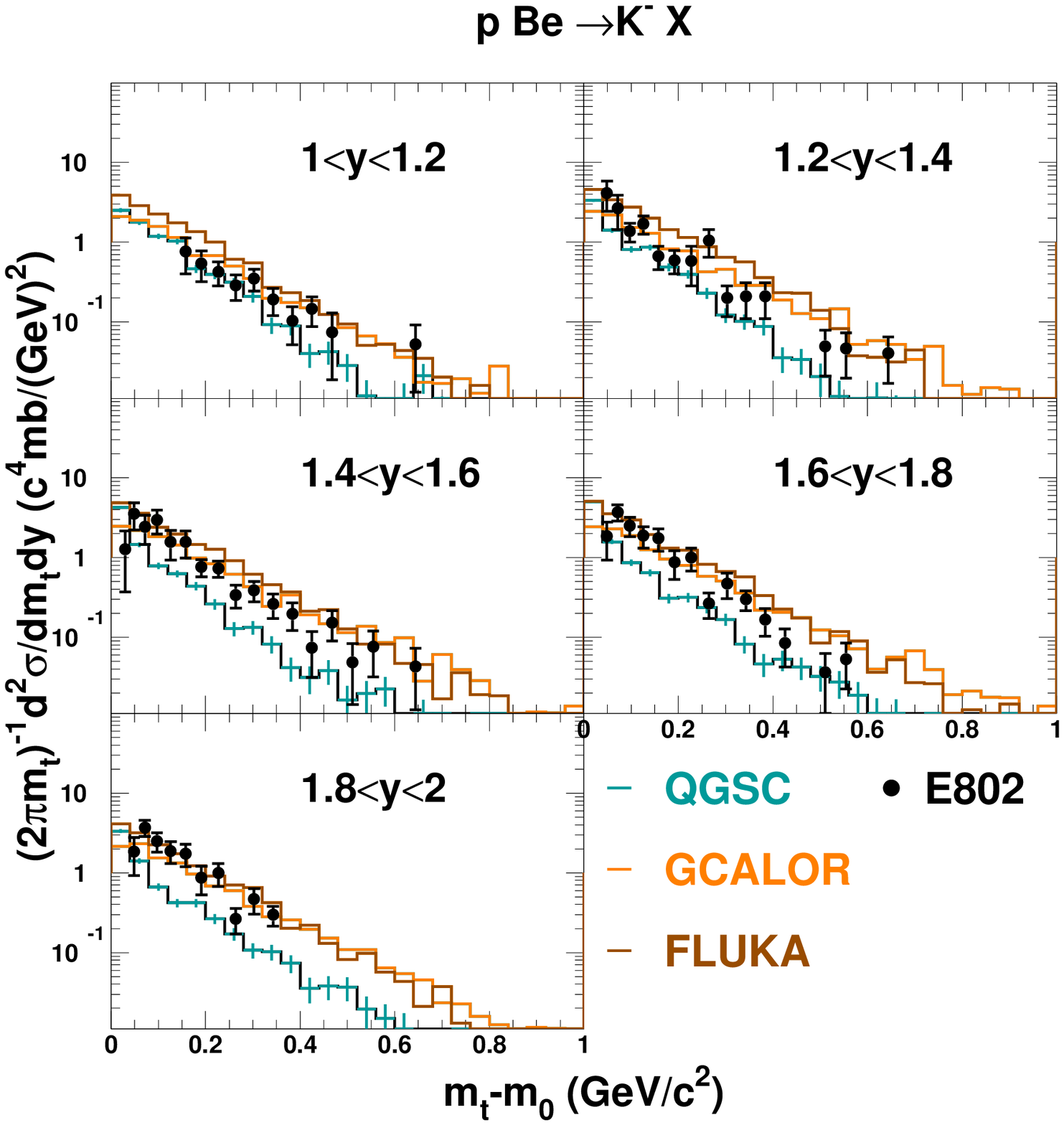,width=.495\linewidth}
\epsfig{file=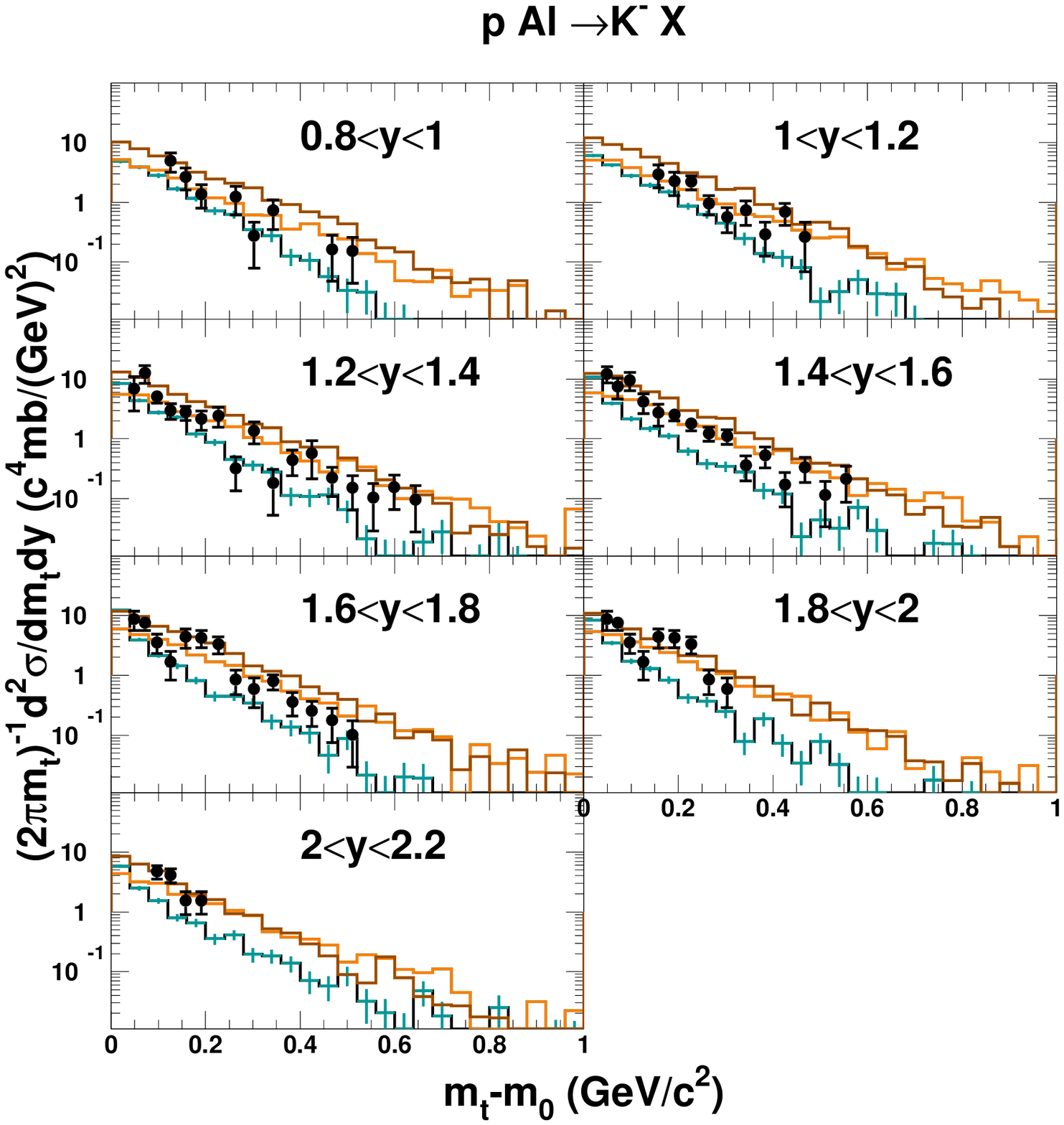,width=.495\linewidth}
\epsfig{file=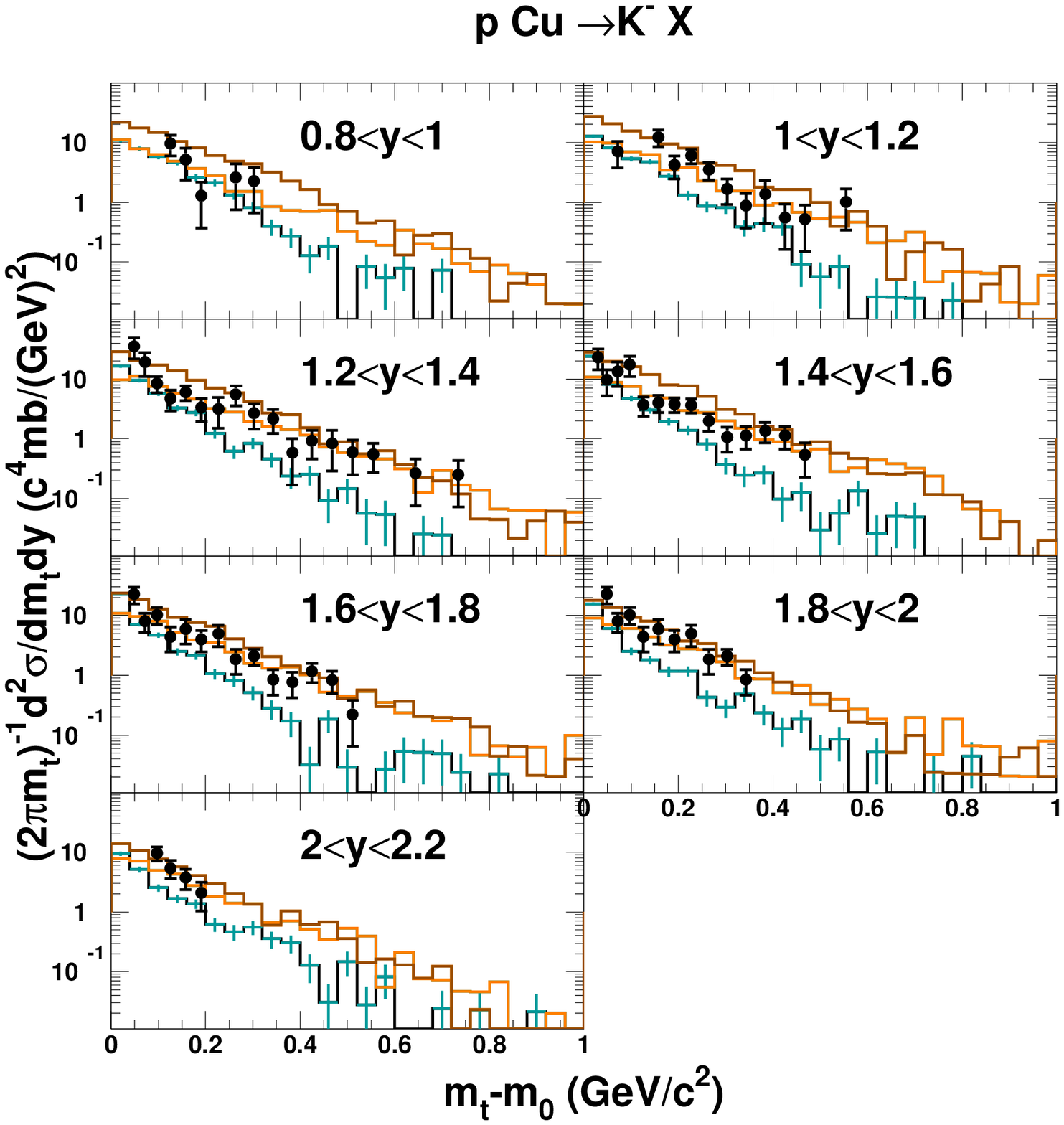,width=.495\linewidth}
\epsfig{file=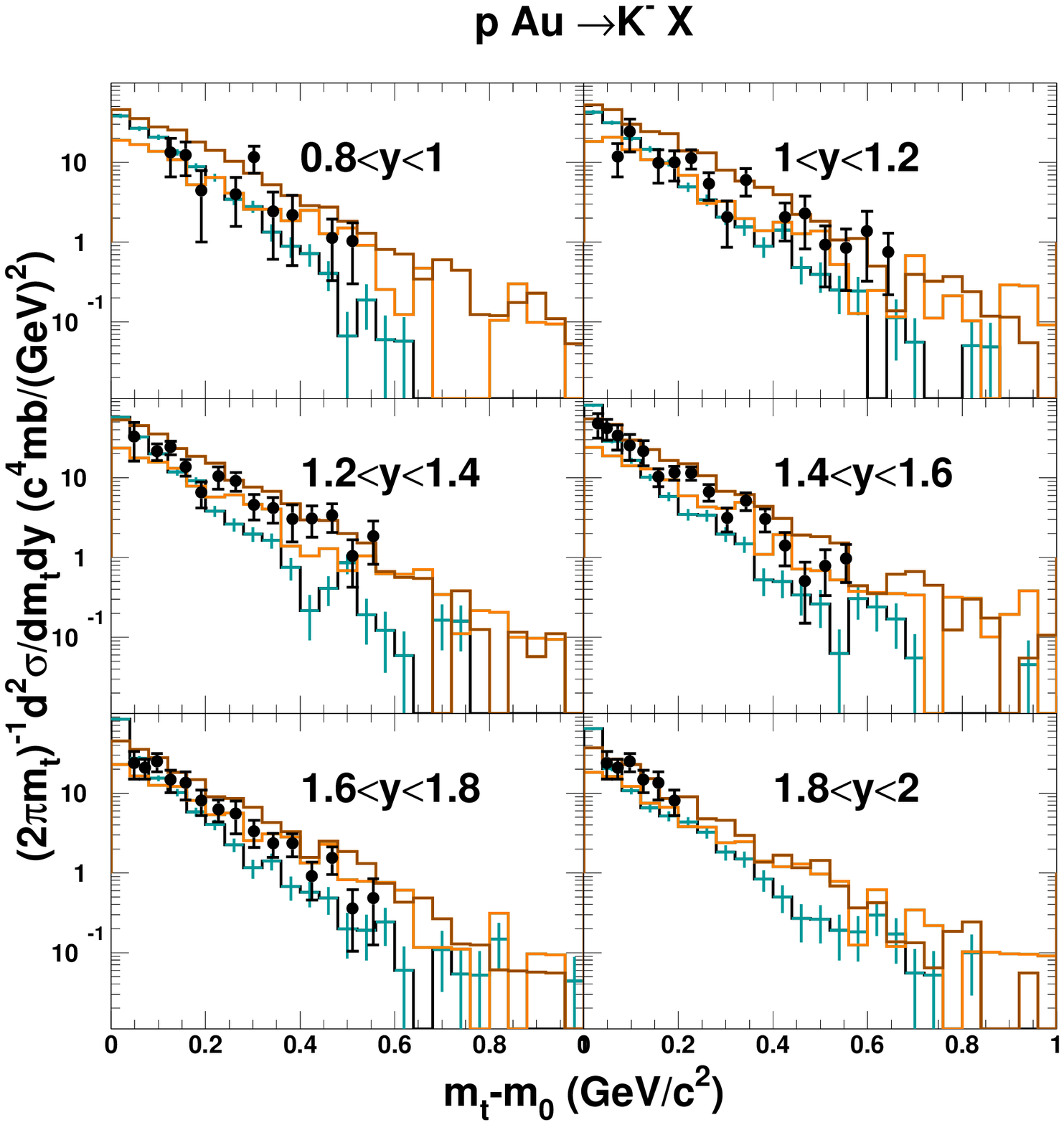,width=.495\linewidth}
\caption{\label{overlay.K-}
The invariant cross section 
$\frac{d^2\sigma}{2\pi m_{\rm t} dm_{\rm t} dy}$ as a function of transverse kinetic energy $m_{\rm t} - m_0$ 
in 0.2 bins of rapidity compared to the simulation results for the 
$K^-$ data
for $p$-Be, $p$-Al, $p$-Cu, and $p$-Au collisions. The statistical uncertainties for the different models is similar and is only shown for QGSC.}
\end{center}\end{figure}

\begin{figure}\begin{center}
\epsfig{file=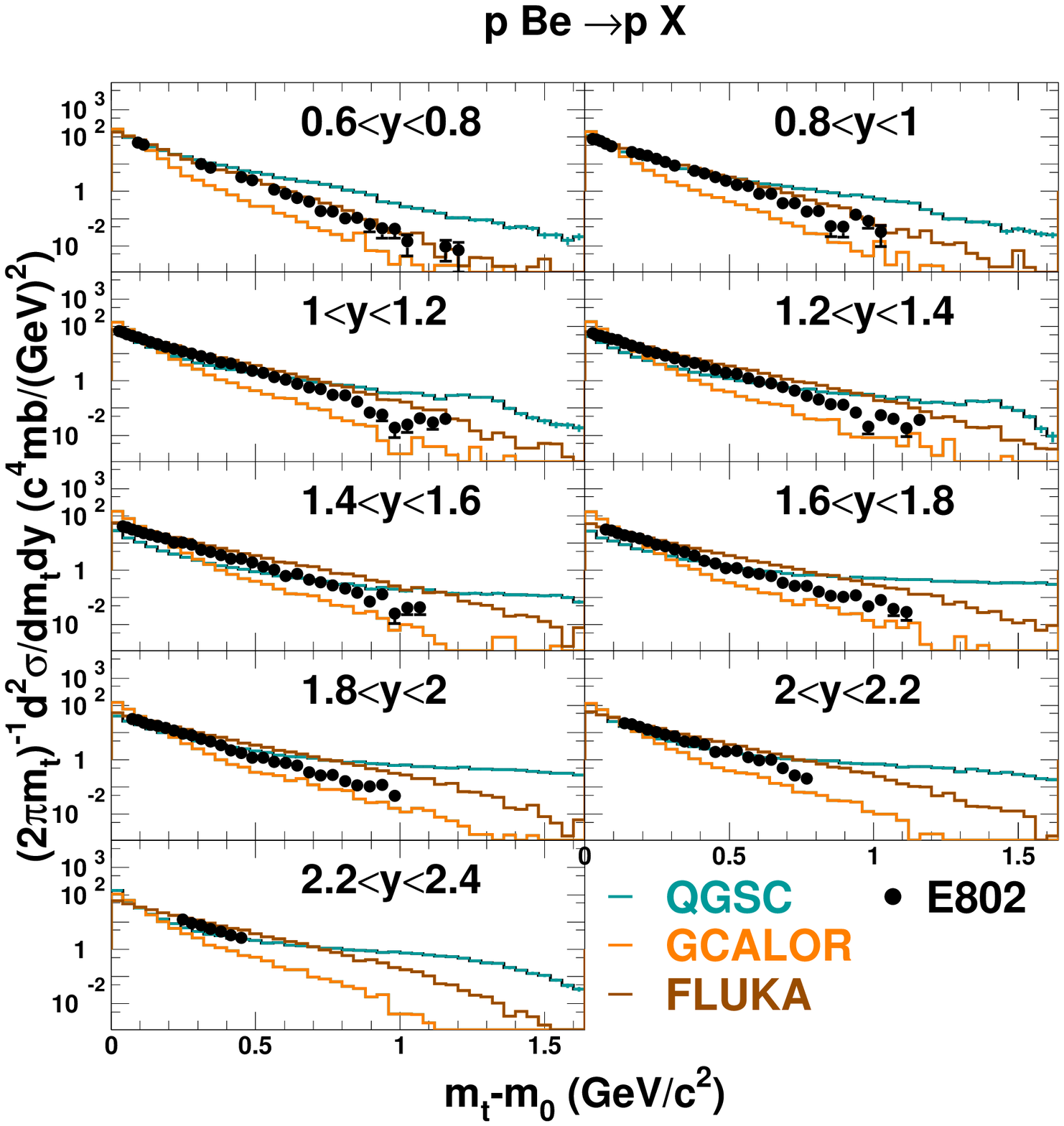,width=.495\linewidth}
\epsfig{file=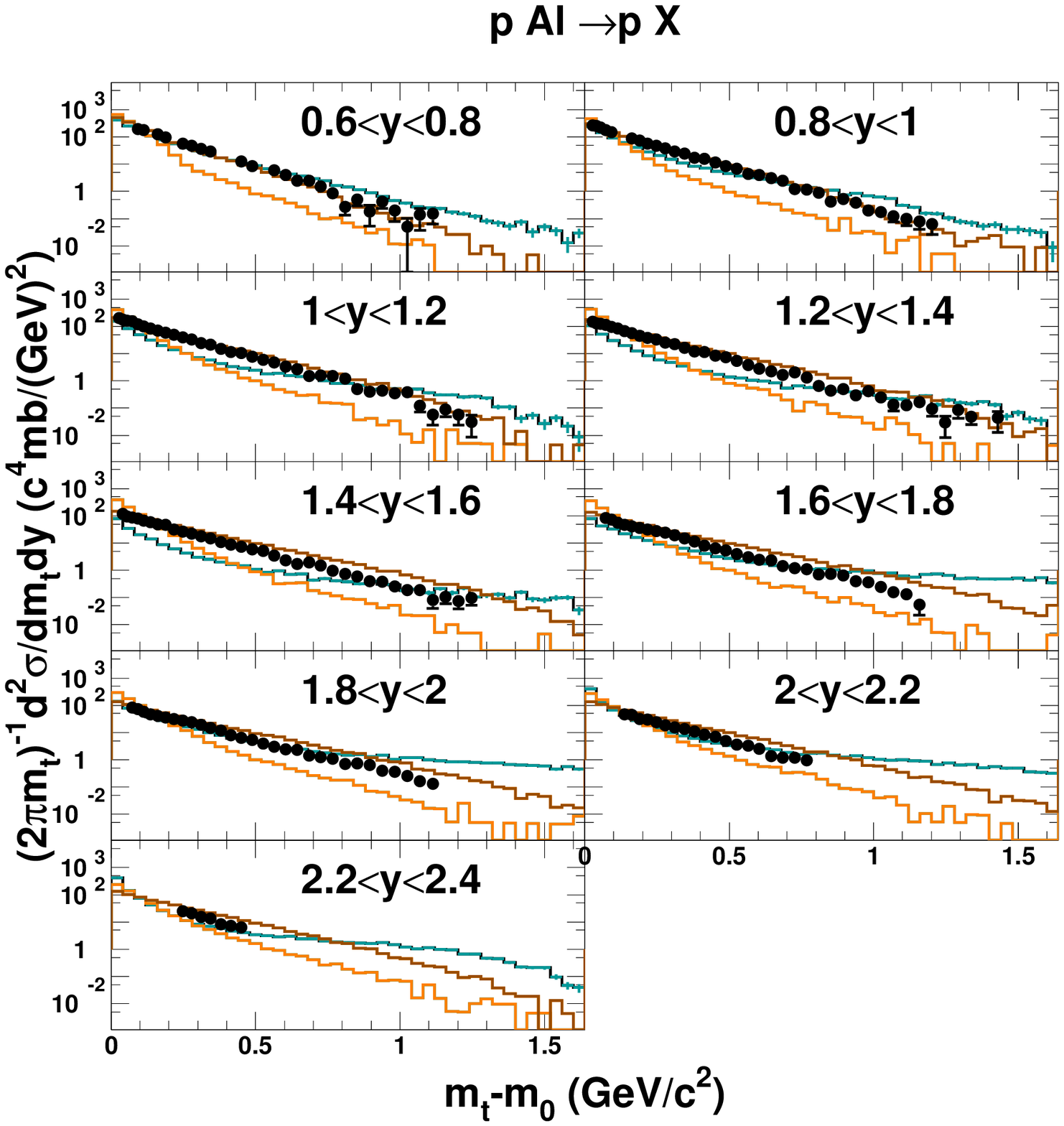,width=.495\linewidth}
\epsfig{file=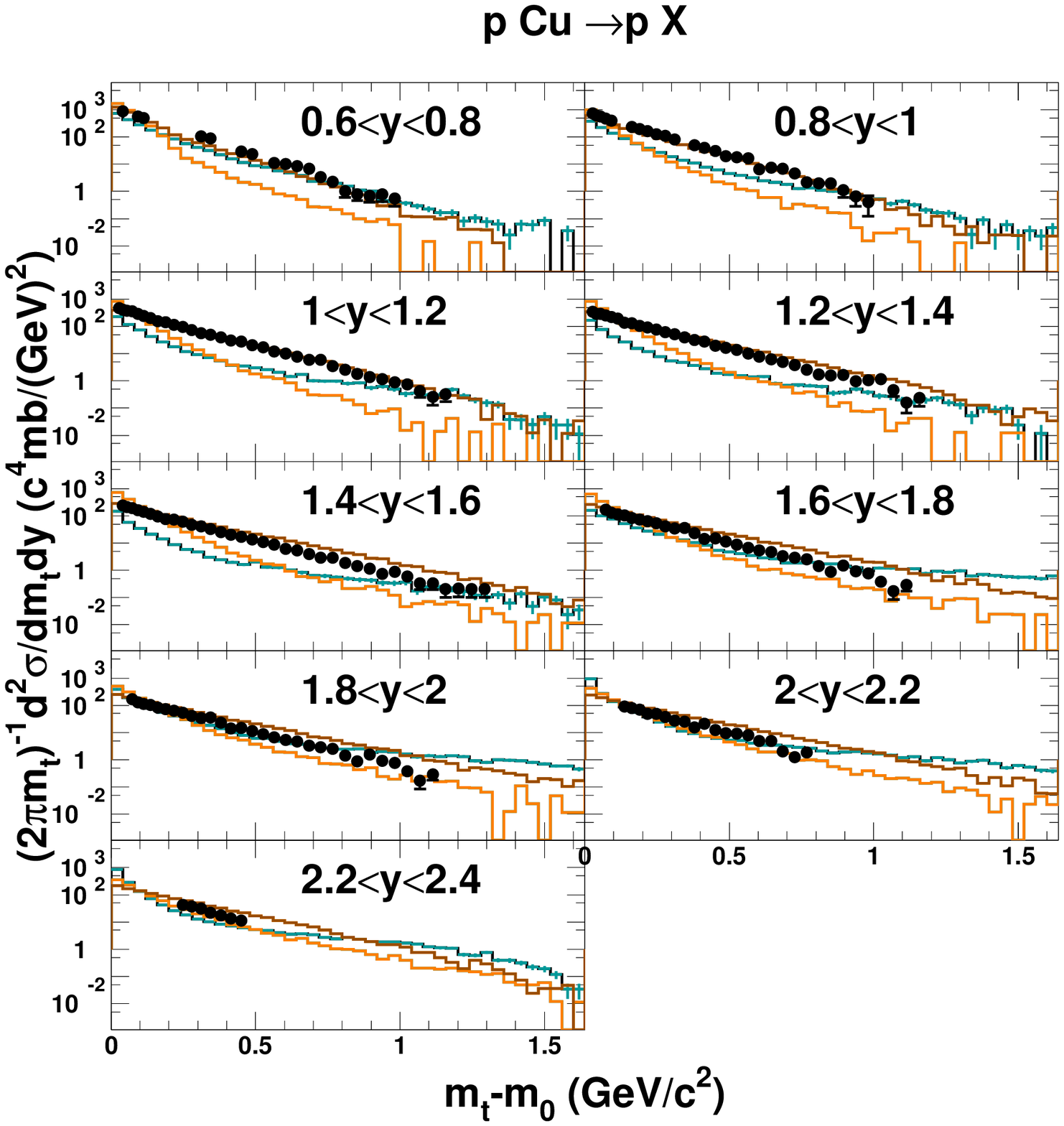,width=.495\linewidth}
\epsfig{file=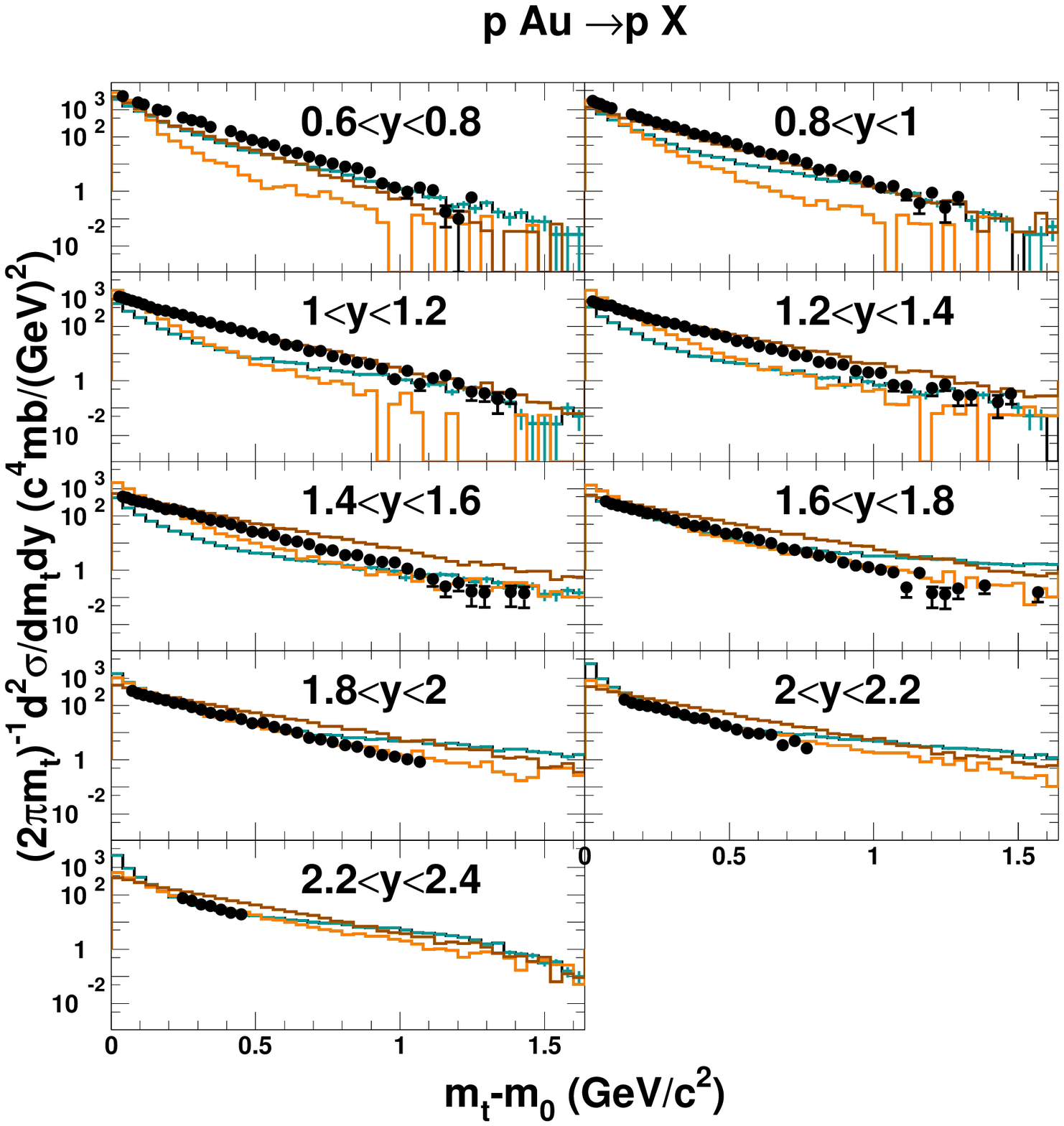,width=.495\linewidth}
\caption{\label{overlay.p}
The invariant cross section 
$\frac{d^2\sigma}{2\pi m_{\rm t} dm_{\rm t} dy}$ as a function of transverse kinetic energy $m_{\rm t} - m_0$ 
in 0.2 bins of rapidity compared to the simulation results for the 
proton data
for $p$-Be, $p$-Al, $p$-Cu, and $p$-Au collisions. The statistical uncertainties for the different models is similar and is only shown for QGSC.}
\end{center}\end{figure}

\begin{figure}\begin{center}
\epsfig{file=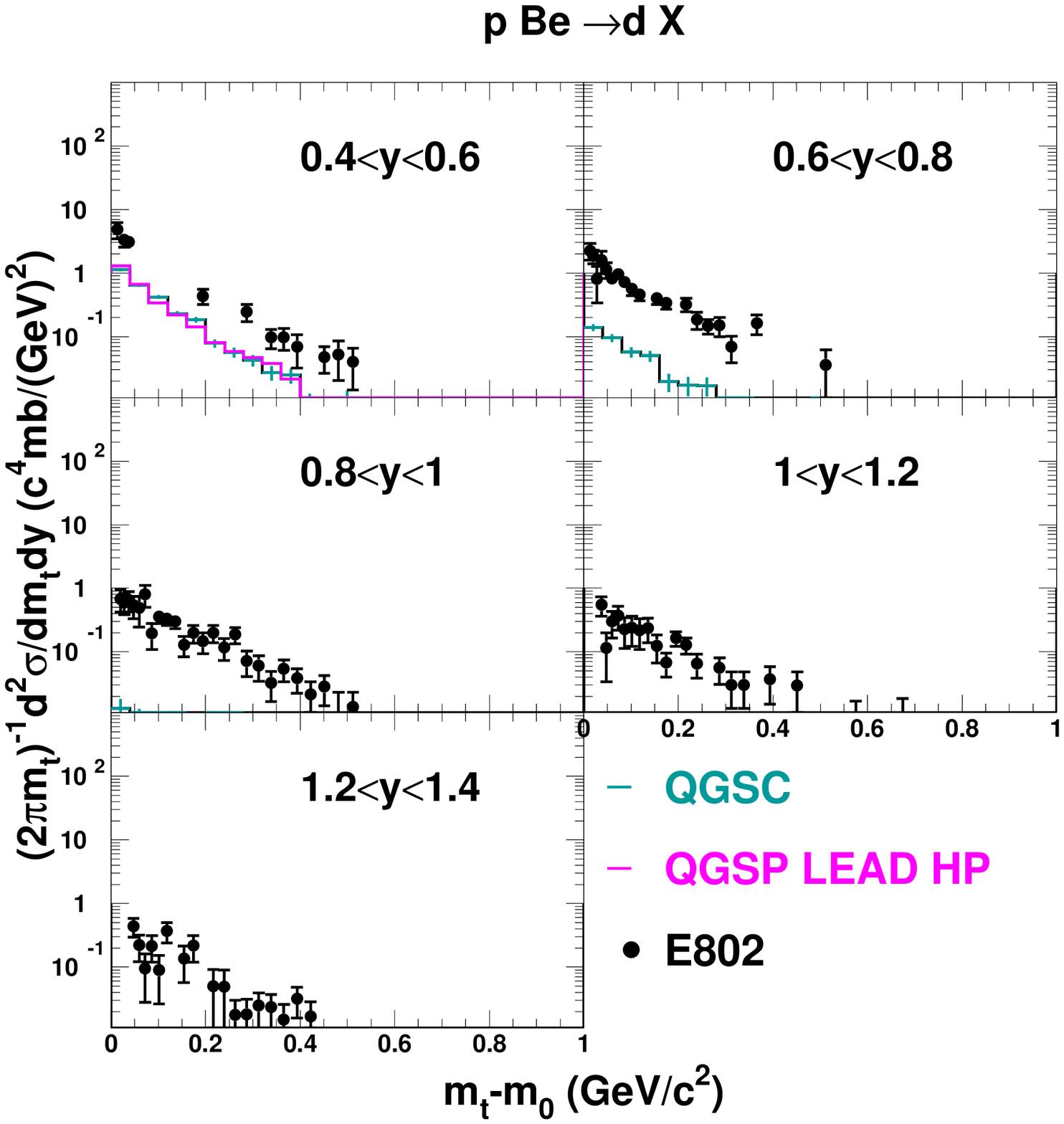,width=.495\linewidth}
\epsfig{file=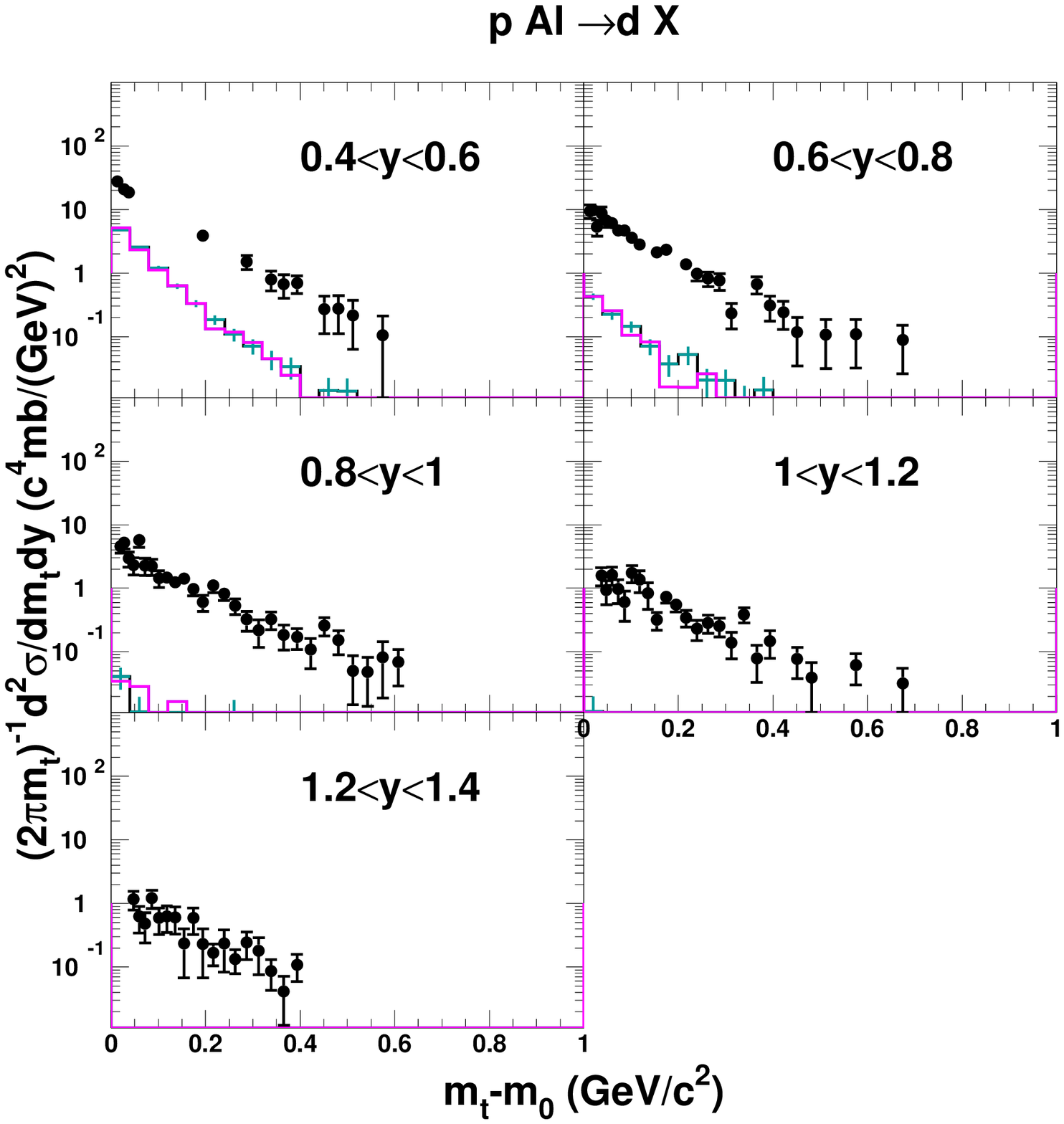,width=.495\linewidth}
\epsfig{file=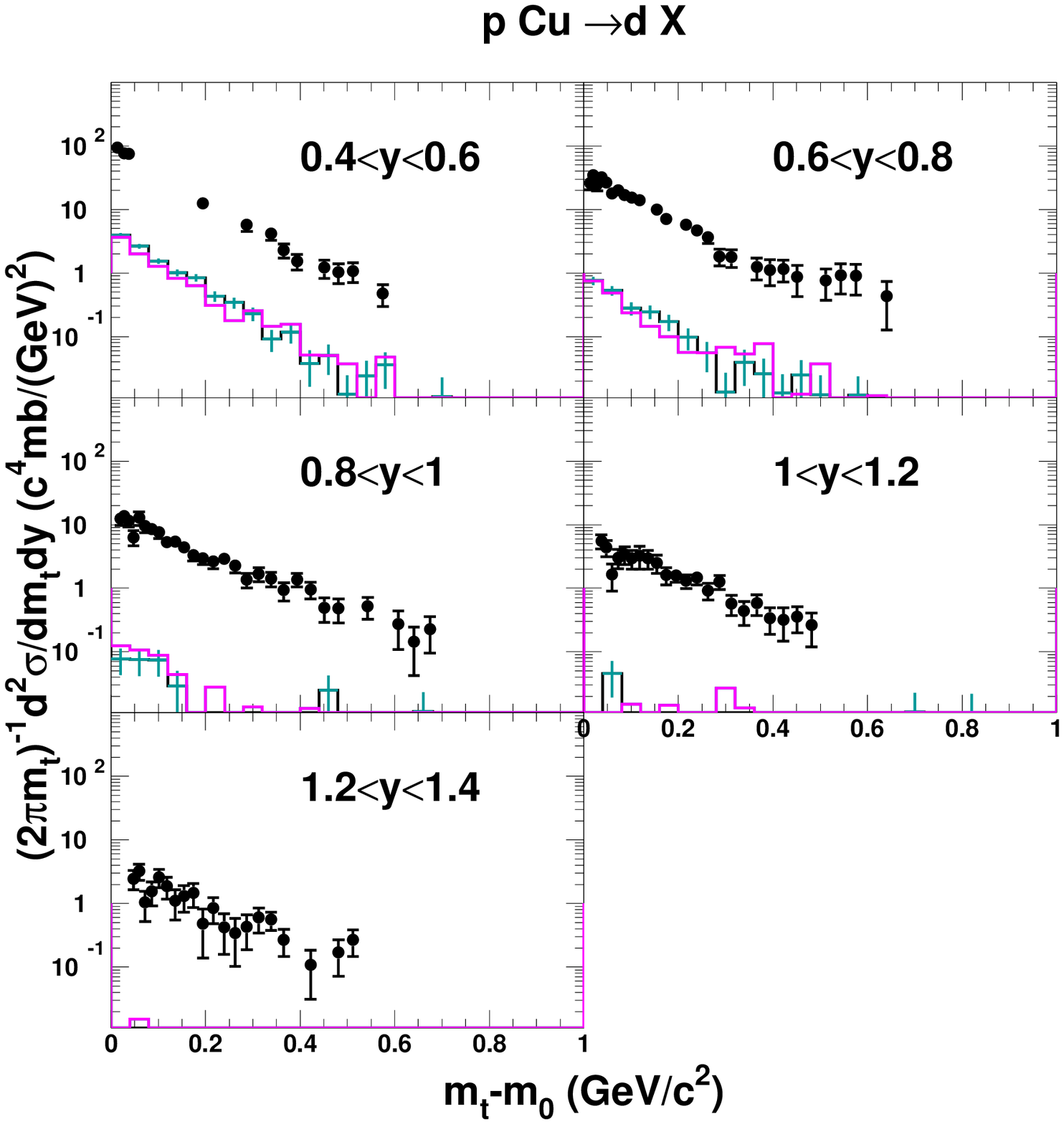,width=.495\linewidth}
\epsfig{file=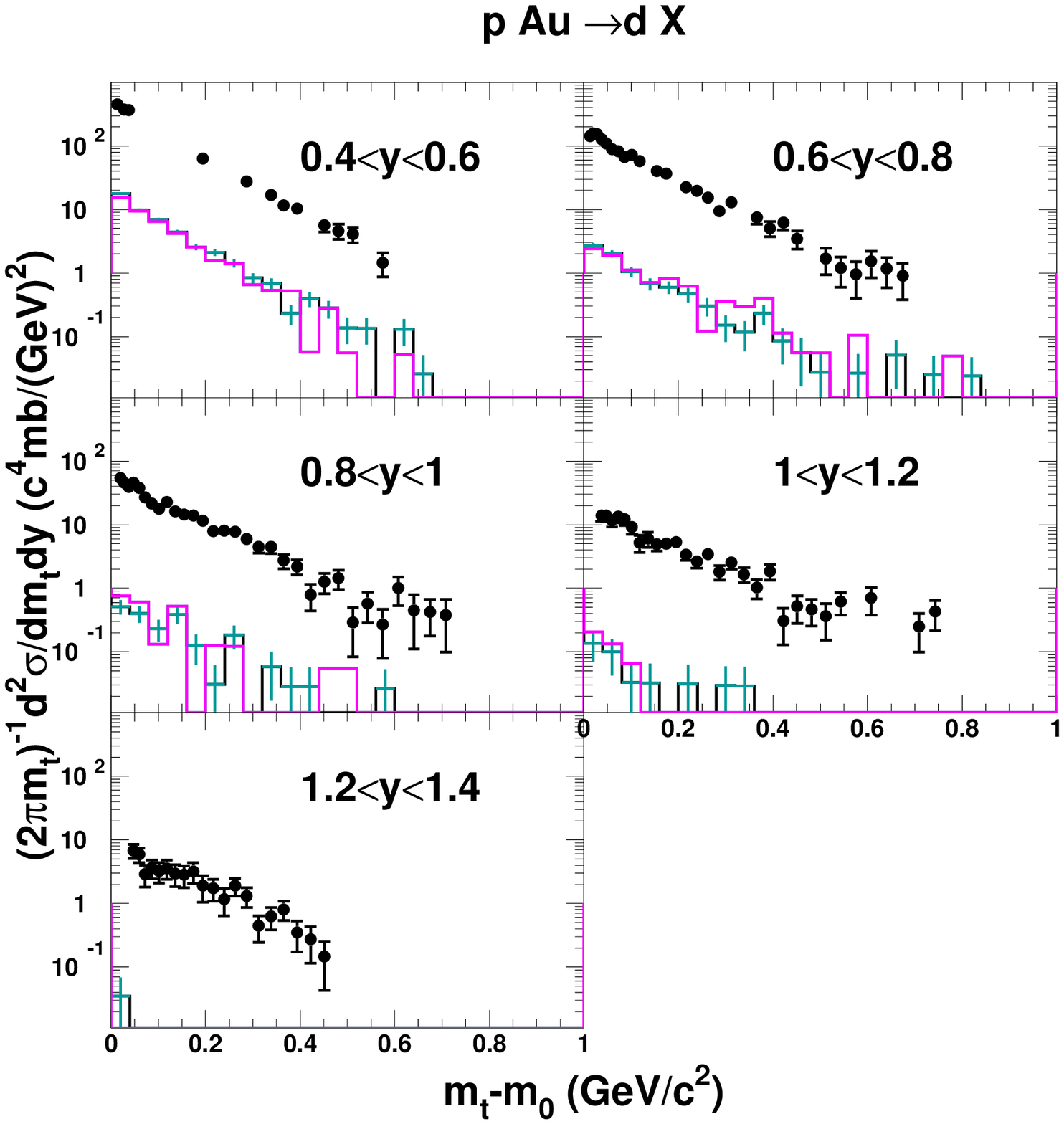,width=.495\linewidth}
\caption{\label{overlay.d}
The invariant cross section 
$\frac{d^2\sigma}{2\pi m_{\rm t} dm_{\rm t} dy}$ as a function of transverse kinetic energy $m_{\rm t} - m_0$ 
in 0.2 bins of rapidity compared to the simulation results for the 
deuteron data
for $p$-Be, $p$-Al, $p$-Cu, and $p$-Au collisions. The statistical uncertainties for the different models is similar and is only shown for QGSC.}
\end{center}\end{figure}

\begin{figure}\begin{center}
\epsfig{file=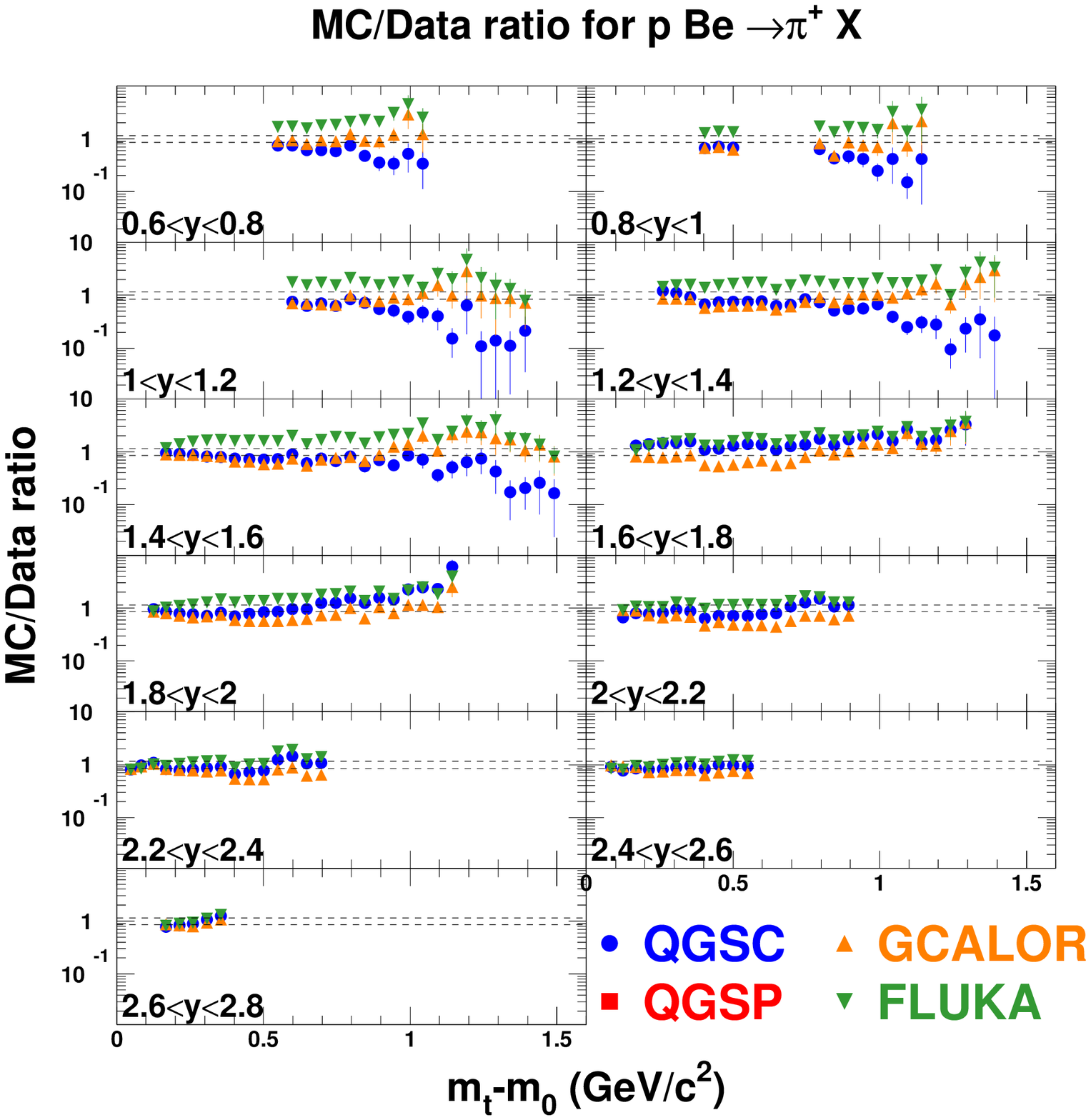,width=.495\linewidth}
\epsfig{file=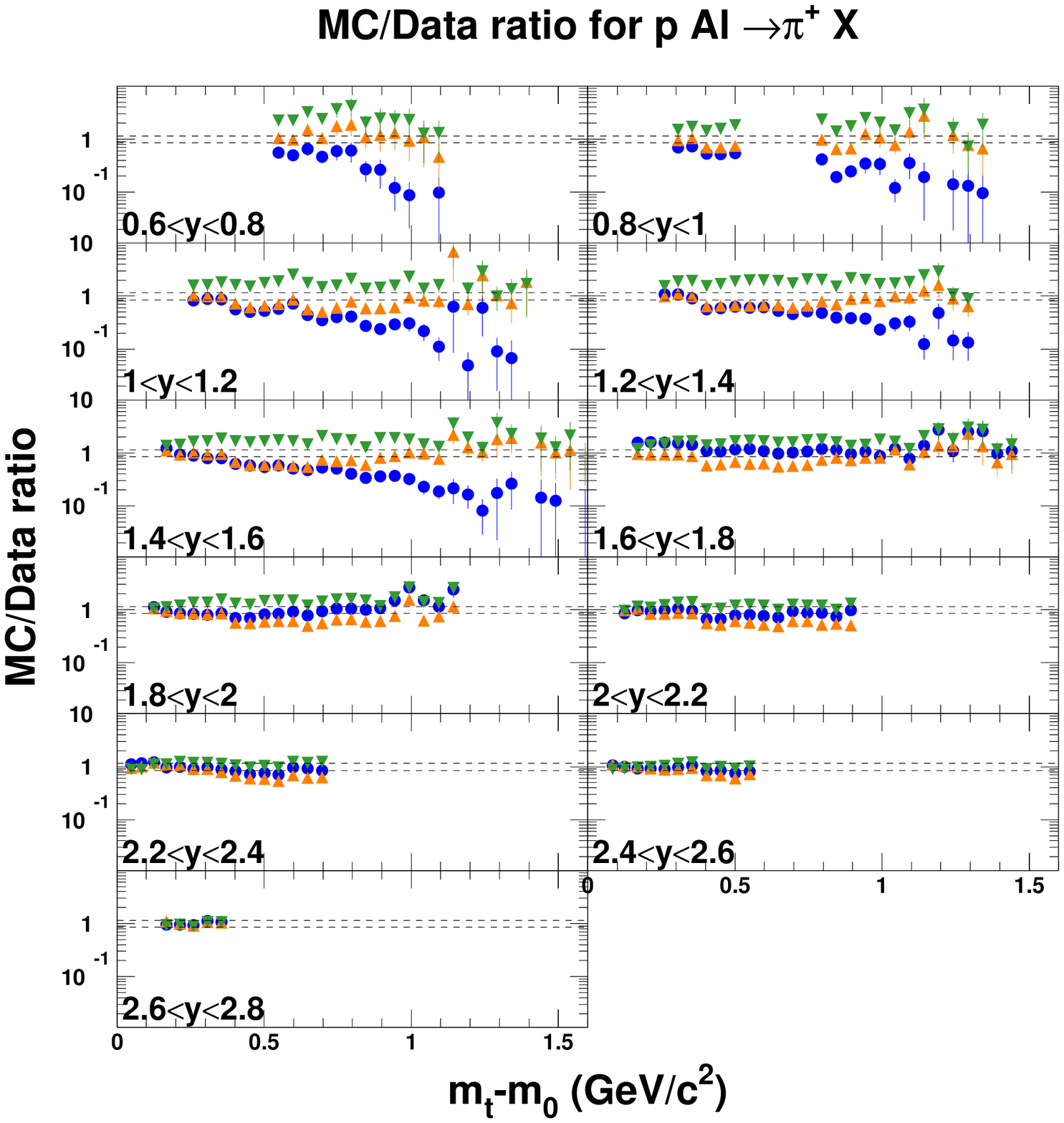,width=.495\linewidth}
\epsfig{file=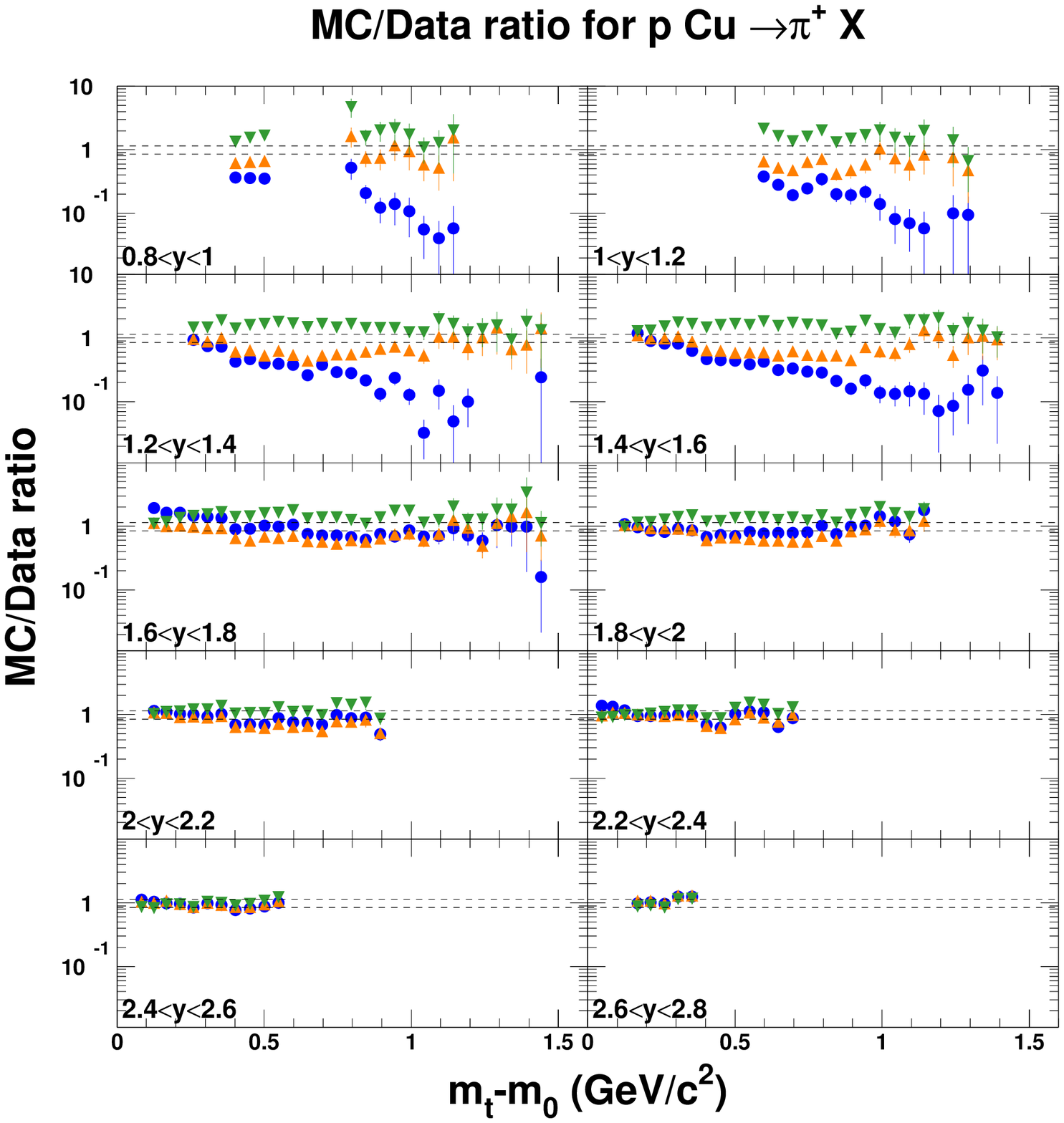,width=.495\linewidth}
\epsfig{file=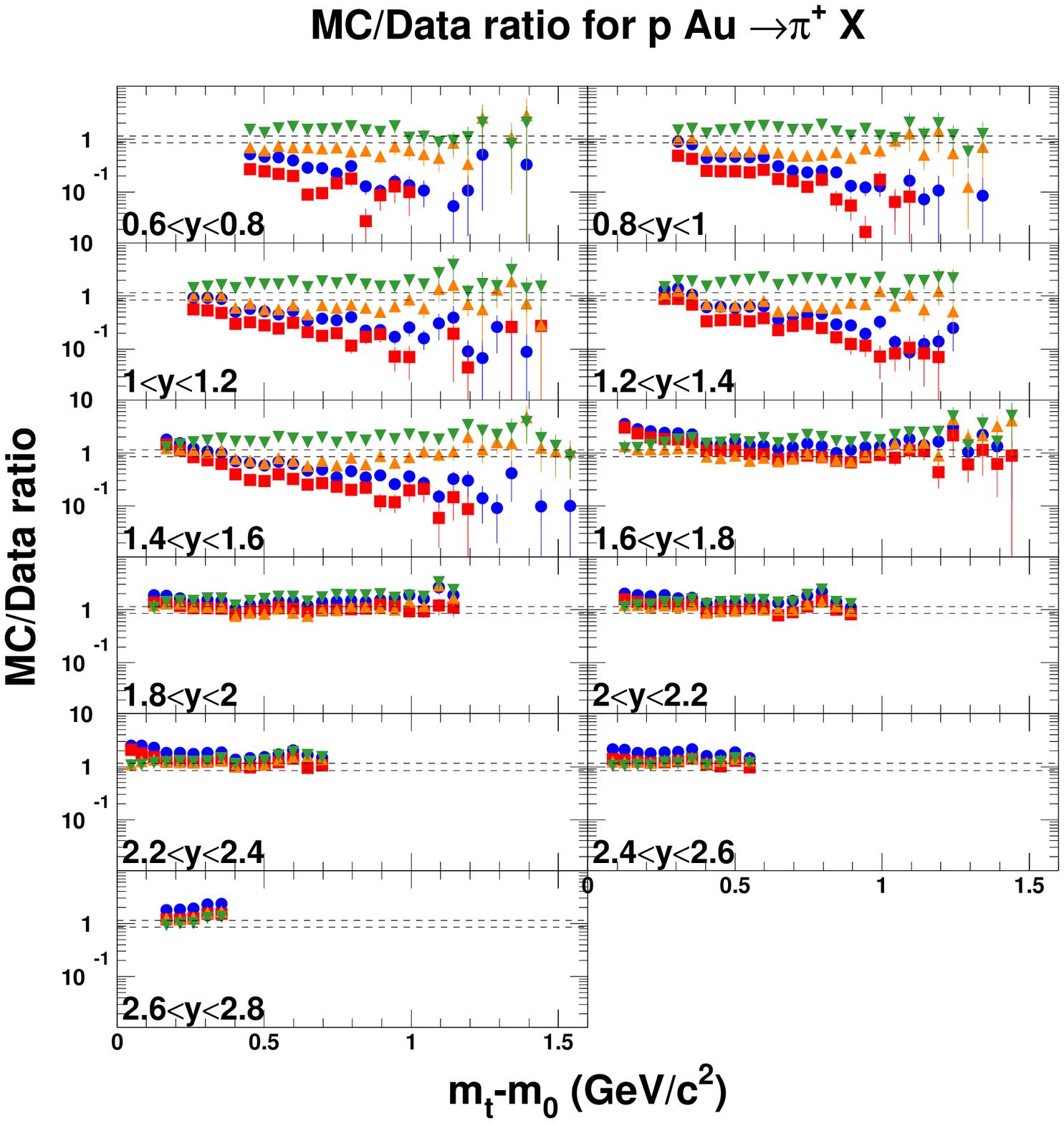,width=.495\linewidth}
\caption{\label{ratio.pi+}
The ratio of the simulated (MC) and data invariant cross sections  
as a function of transverse kinetic energy $m_{\rm t} - m_0$ 
in 0.2 bins of rapidity for the  
$\pi^+$ data
for $p$-Be, $p$-Al, $p$-Cu, and $p$-Au collisions.
The horizontal dashed lines indicate the $\pm 15\%$ normalization uncertainty of the E802 data. 
Only the QGSC ratios are shown as all the GEANT4 models given consistent predictions 
except for the gold target.}
\end{center}\end{figure}

\begin{figure}\begin{center}
\epsfig{file=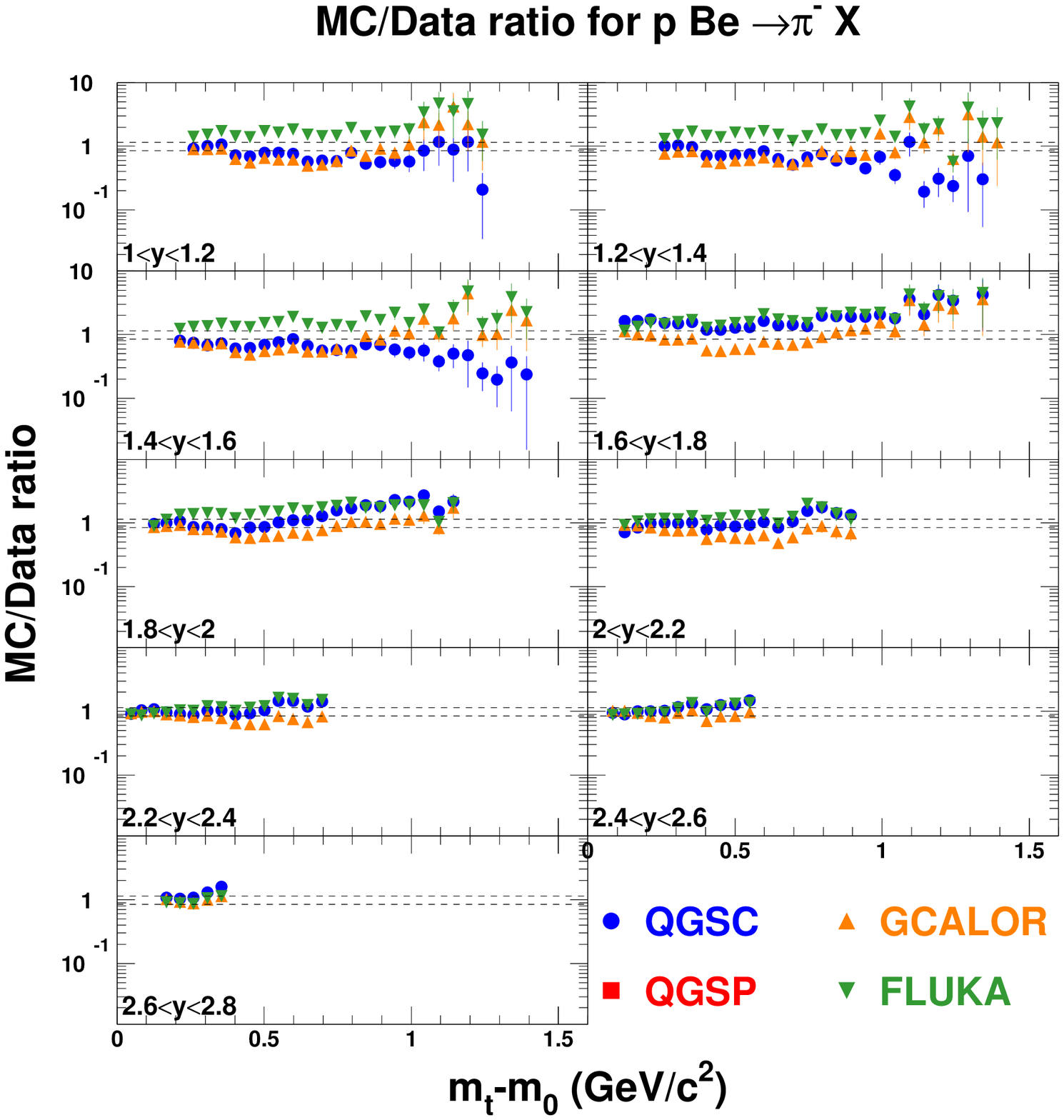,width=.495\linewidth}
\epsfig{file=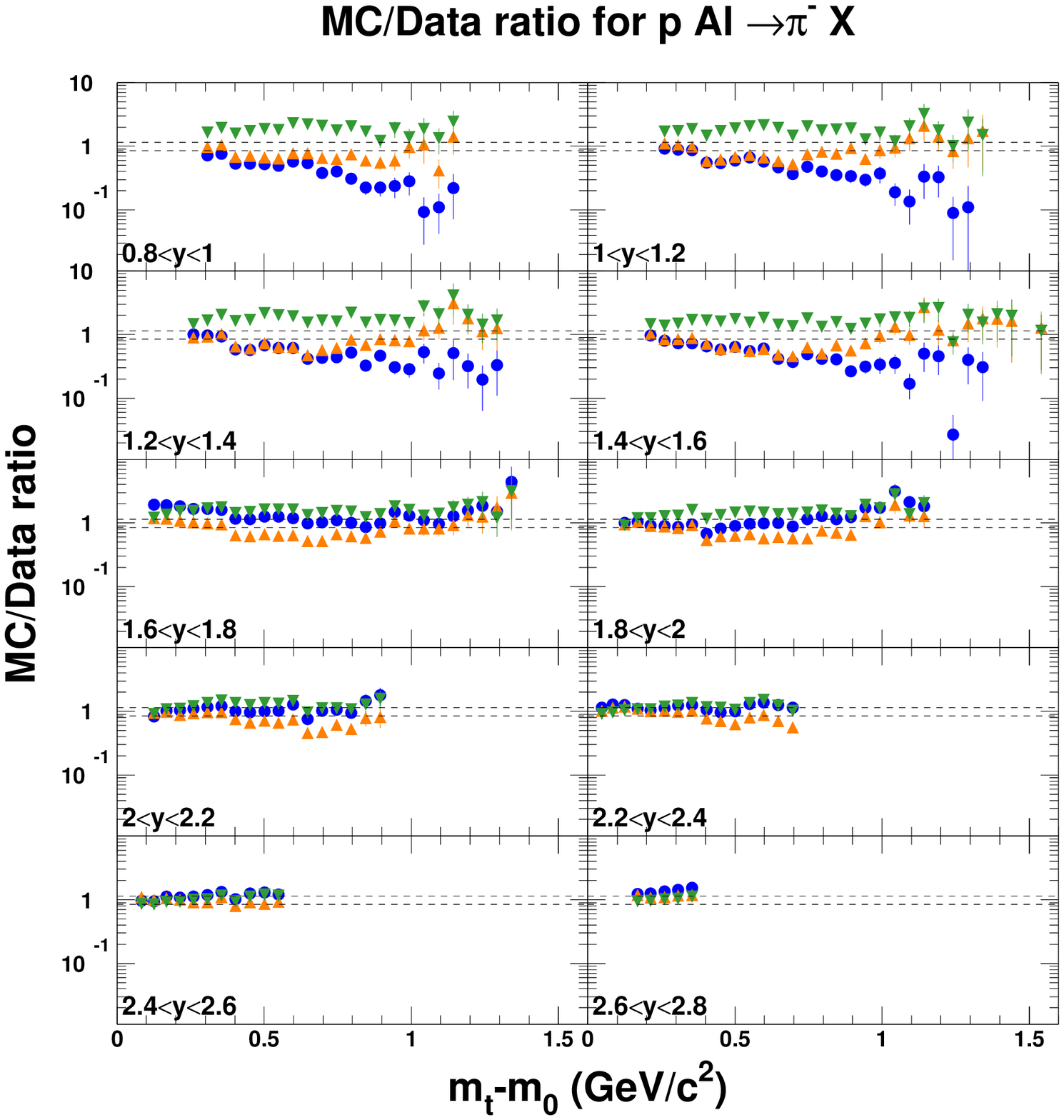,width=.495\linewidth}
\epsfig{file=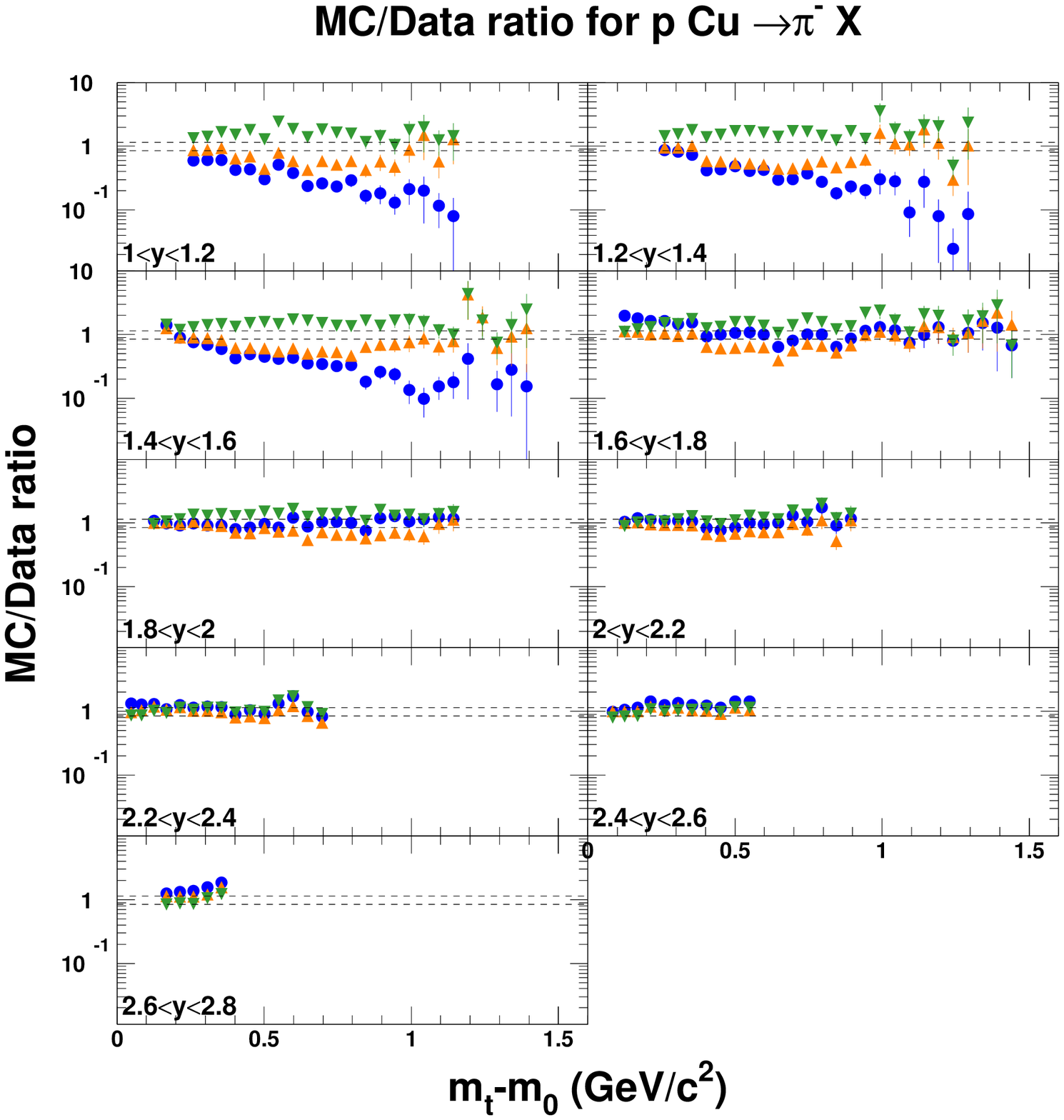,width=.495\linewidth}
\epsfig{file=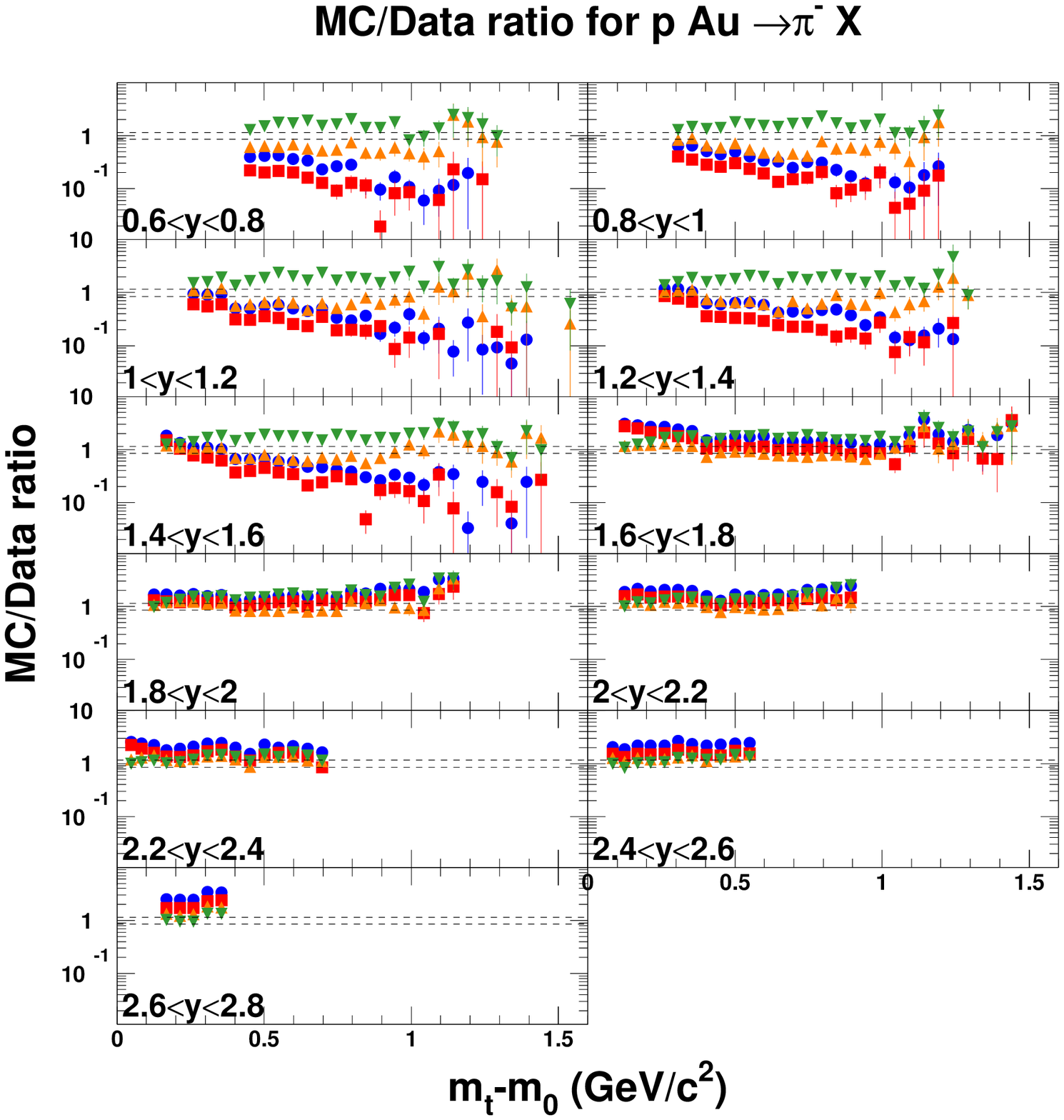,width=.495\linewidth}
\caption{\label{ratio.pi-}
The ratio of the simulated (MC) and data invariant cross sections  
as a function of transverse kinetic energy $m_{\rm t} - m_0$ 
in 0.2 bins of rapidity for the  
$\pi^-$ data
for $p$-Be, $p$-Al, $p$-Cu, and $p$-Au collisions.
The horizontal dashed lines indicate the $\pm 15\%$ normalization uncertainty of the E802 data. 
Only the QGSC ratios are shown as all the GEANT4 models given consistent predictions 
except for the gold target.}
\end{center}\end{figure}

\begin{figure}\begin{center}
\epsfig{file=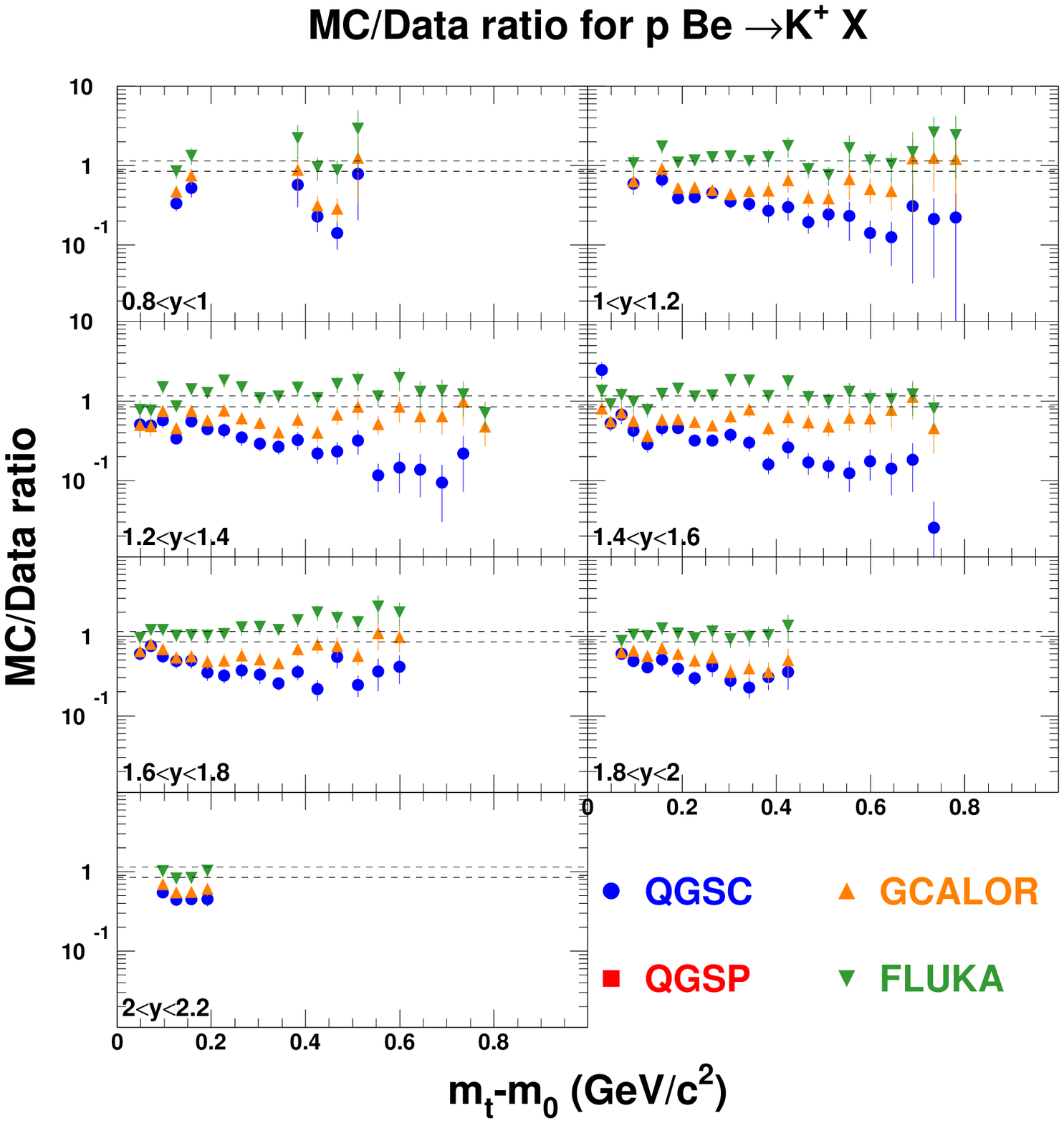,width=.495\linewidth}
\epsfig{file=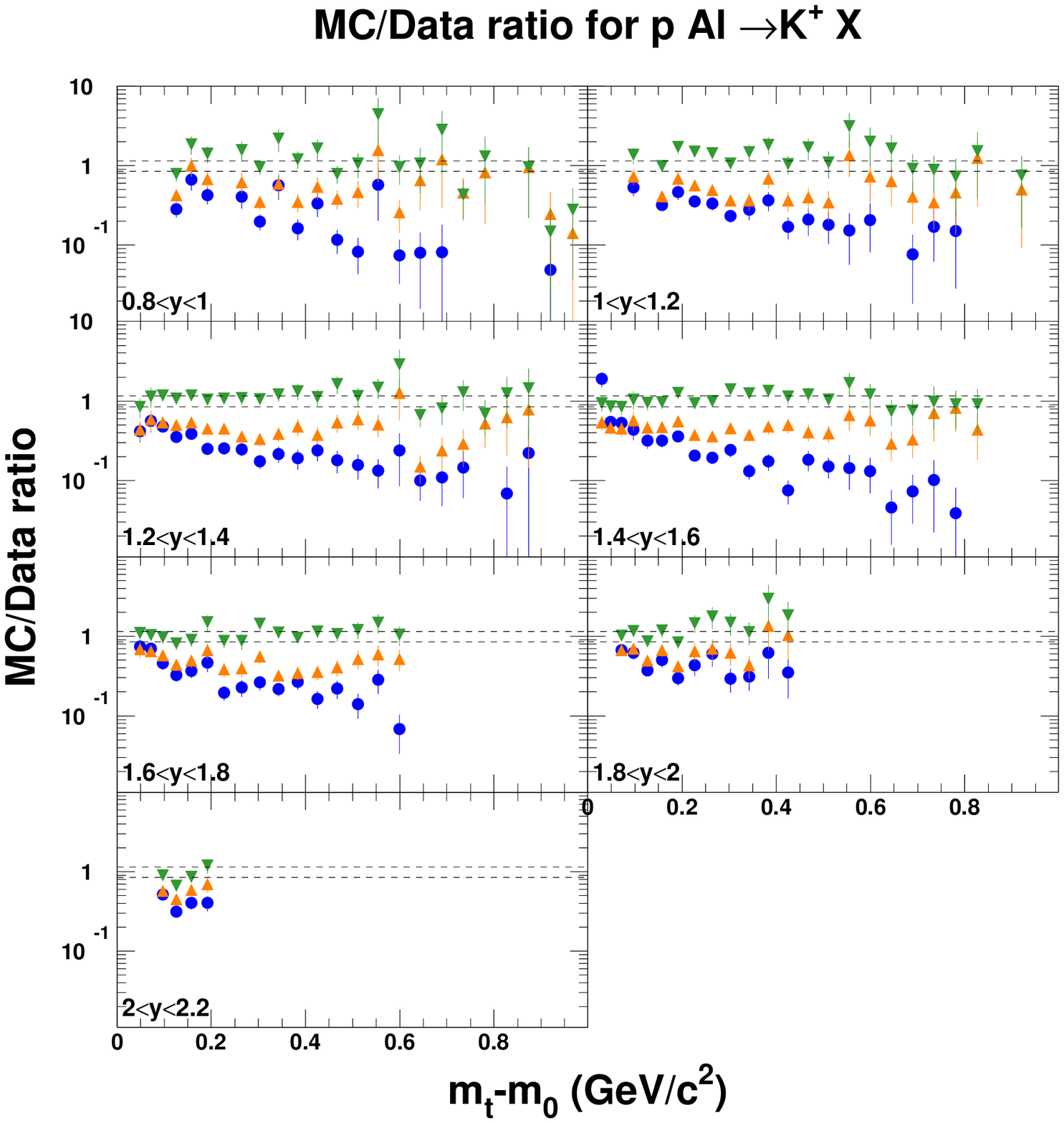,width=.495\linewidth}
\epsfig{file=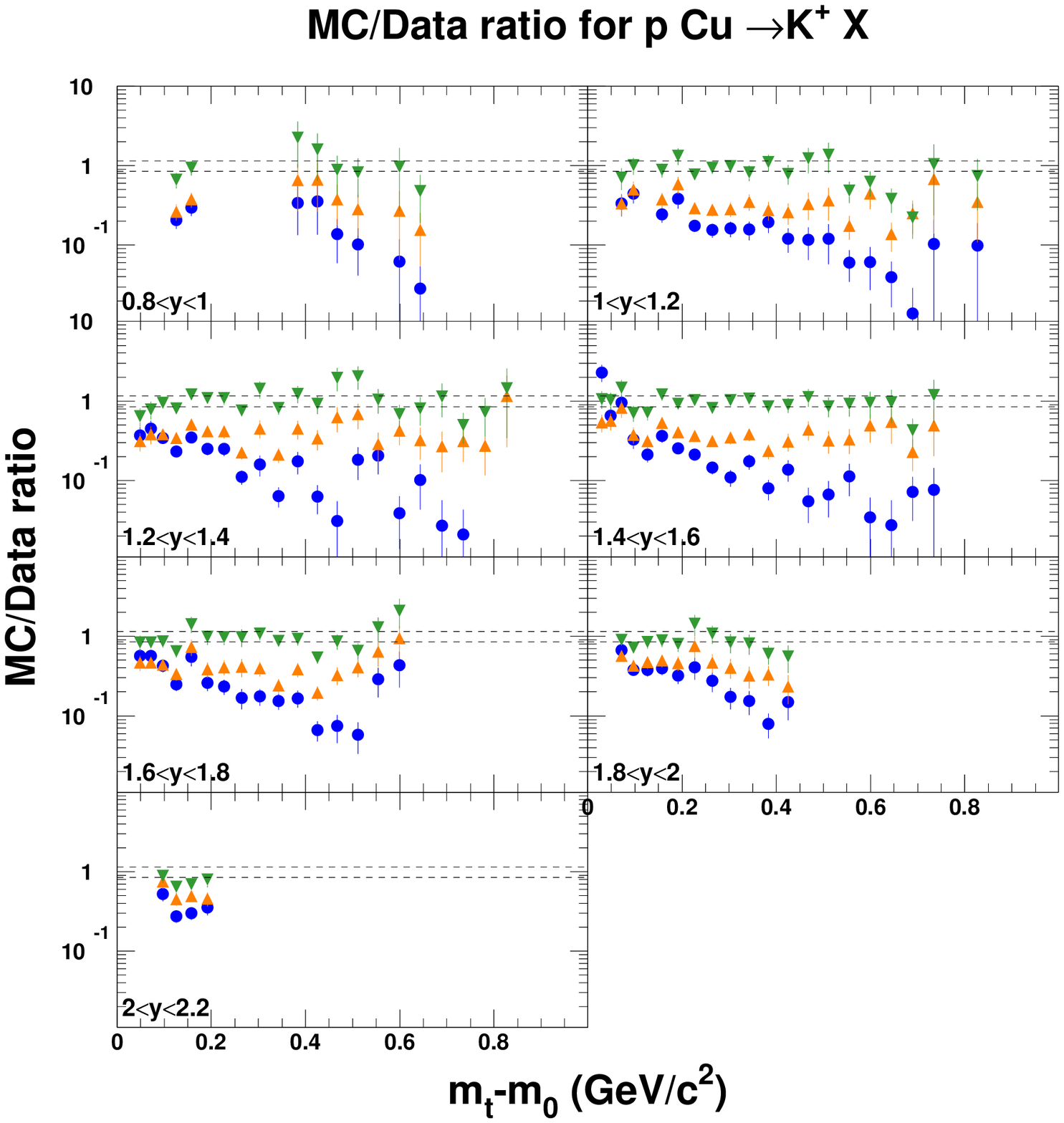,width=.495\linewidth}
\epsfig{file=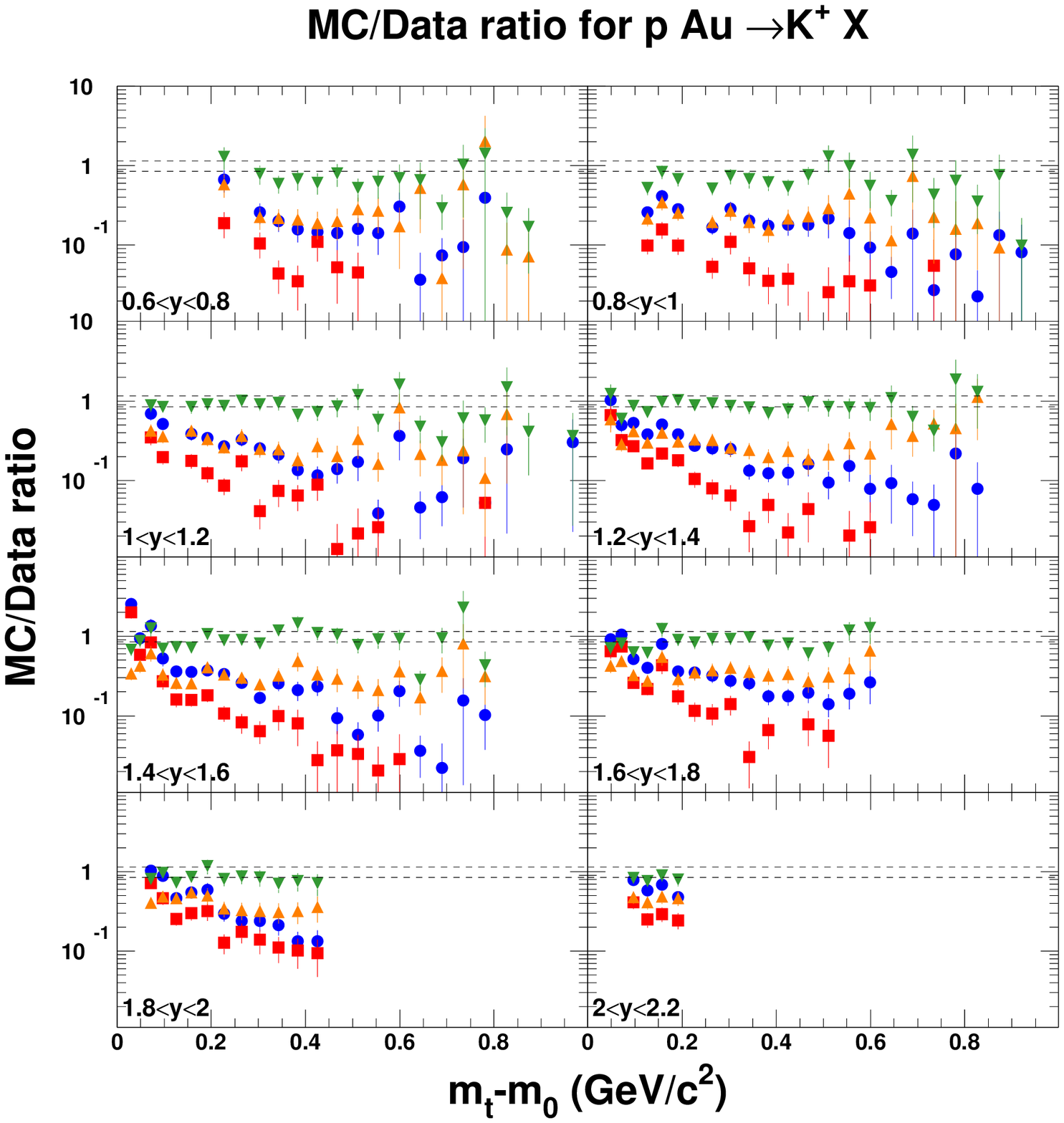,width=.495\linewidth}
\caption{\label{ratio.K+}
The ratio of the simulated (MC) and data invariant cross sections  
as a function of transverse kinetic energy $m_{\rm t} - m_0$ 
in 0.2 bins of rapidity for the  
$K^+$ data
for $p$-Be, $p$-Al, $p$-Cu, and $p$-Au collisions.
The horizontal dashed lines indicate the $\pm 15\%$ normalization uncertainty of the E802 data. 
Only the QGSC ratios are shown as all the GEANT4 models given consistent predictions 
except for the gold target.}
\end{center}\end{figure}

\begin{figure}\begin{center}
\epsfig{file=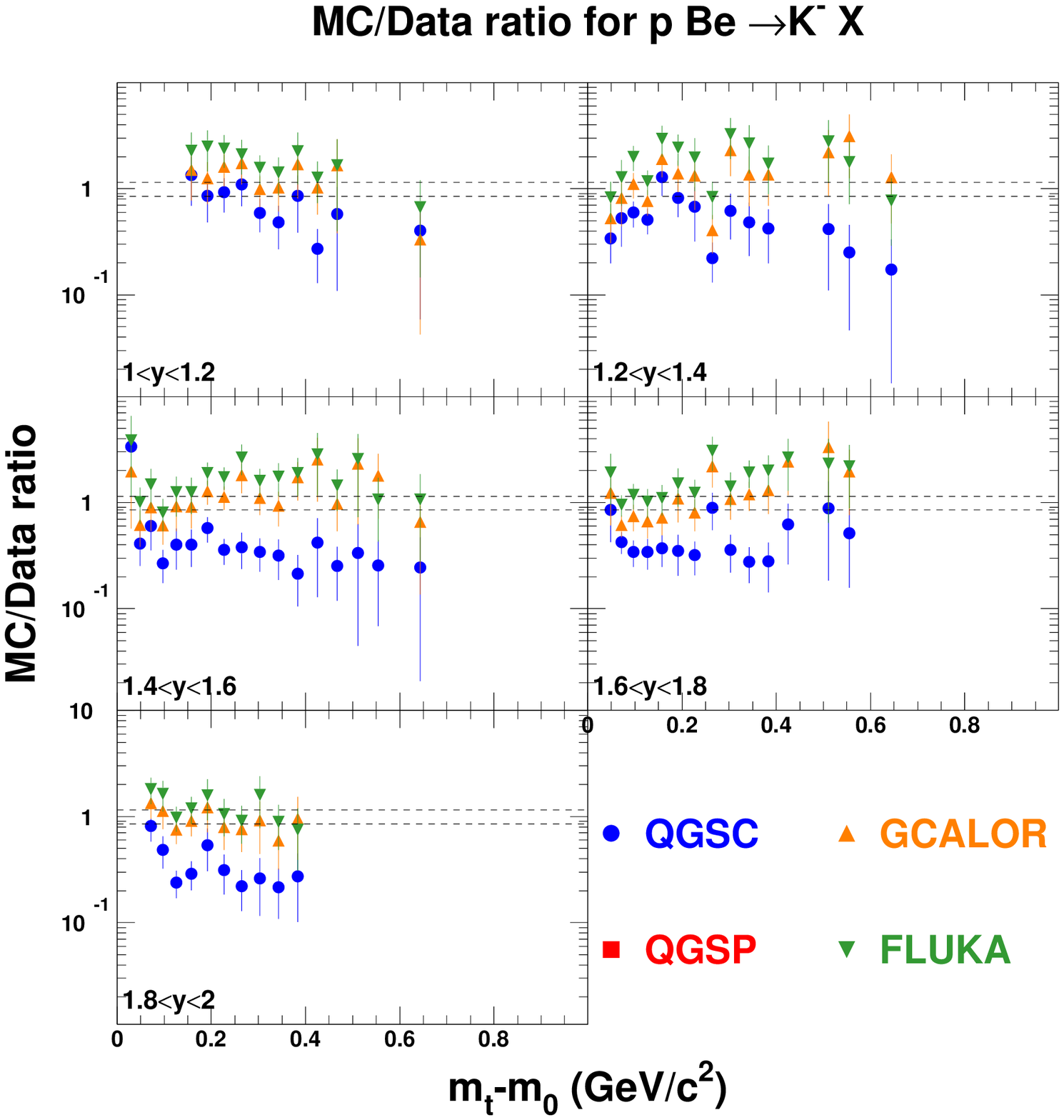,width=.495\linewidth}
\epsfig{file=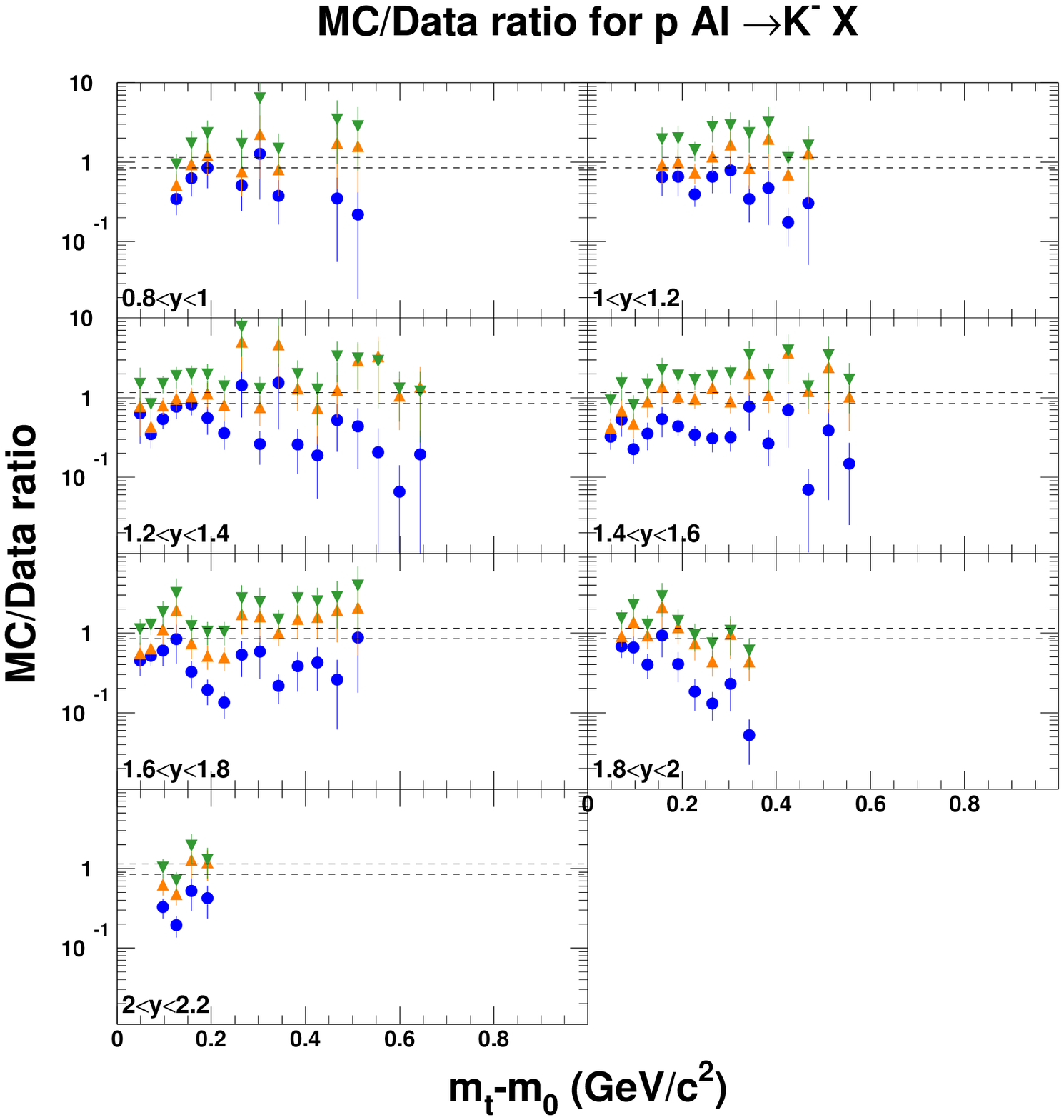,width=.495\linewidth}
\epsfig{file=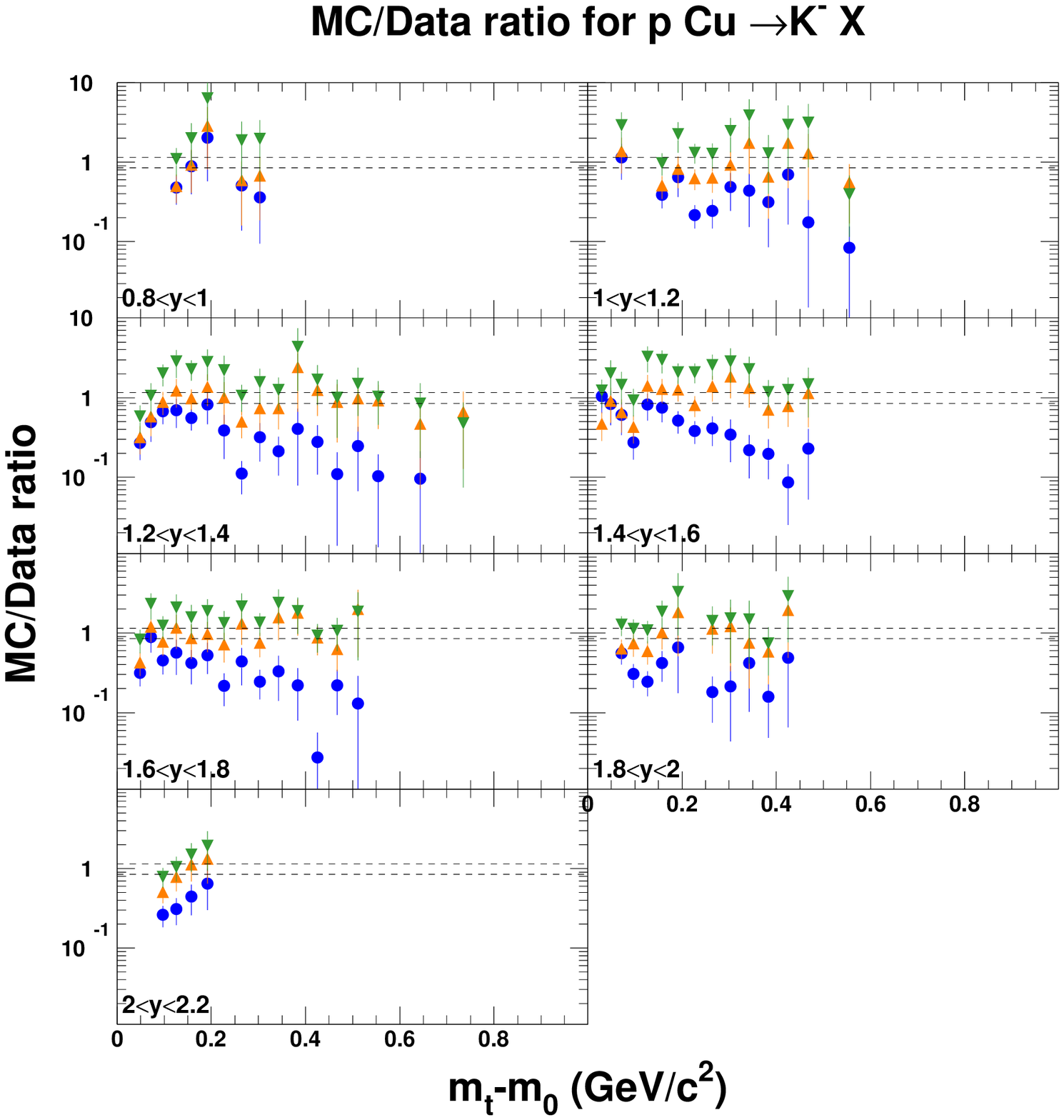,width=.495\linewidth}
\epsfig{file=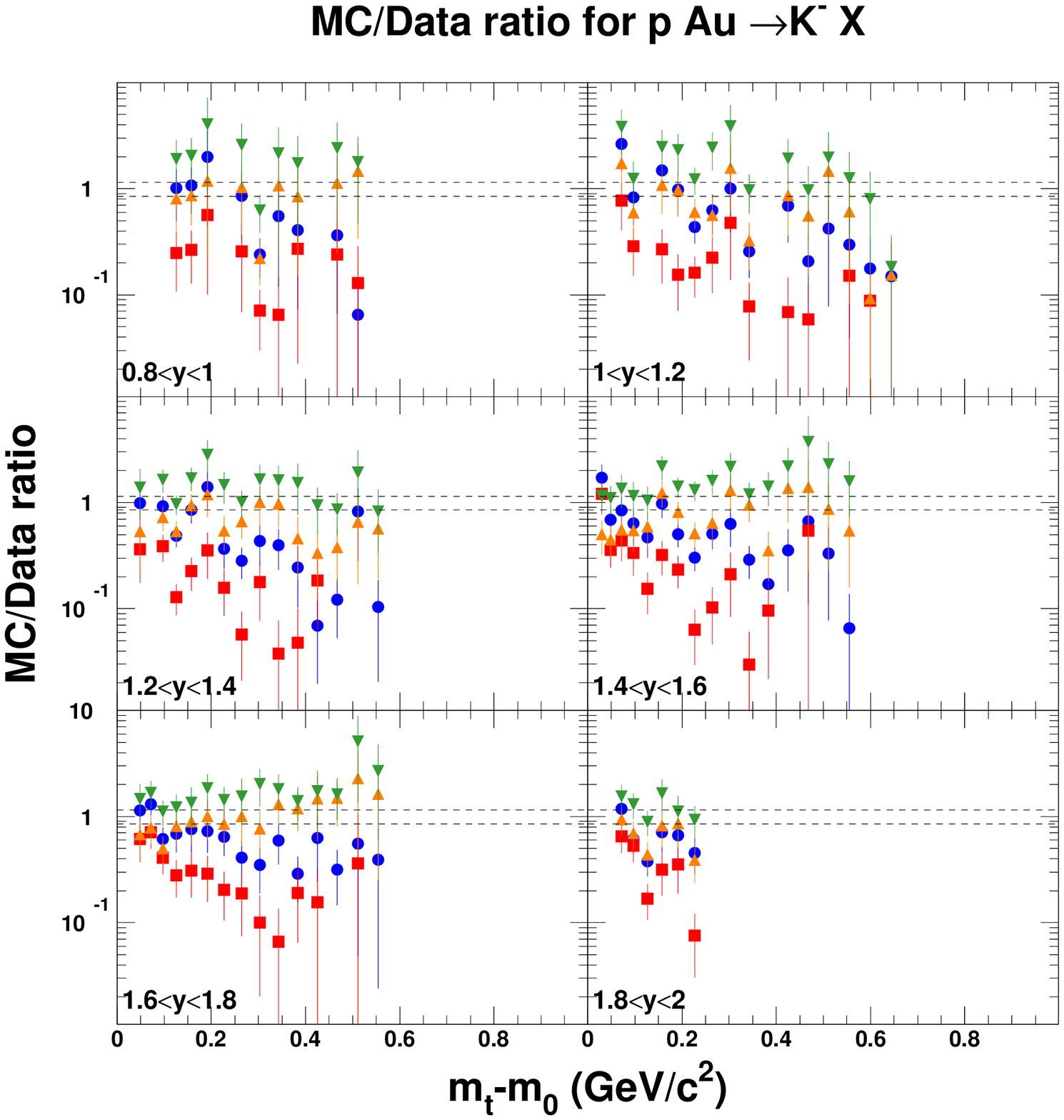,width=.495\linewidth}
\caption{\label{ratio.K-}
The ratio of the simulated (MC) and data invariant cross sections  
as a function of transverse kinetic energy $m_{\rm t} - m_0$ 
in 0.2 bins of rapidity for the  
$K^-$ data
for $p$-Be, $p$-Al, $p$-Cu, and $p$-Au collisions.
The horizontal dashed lines indicate the $\pm 15\%$ normalization uncertainty of the E802 data. 
Only the QGSC ratios are shown as all the GEANT4 models given consistent predictions 
except for the gold target.}
\end{center}\end{figure}

\begin{figure}\begin{center}
\epsfig{file=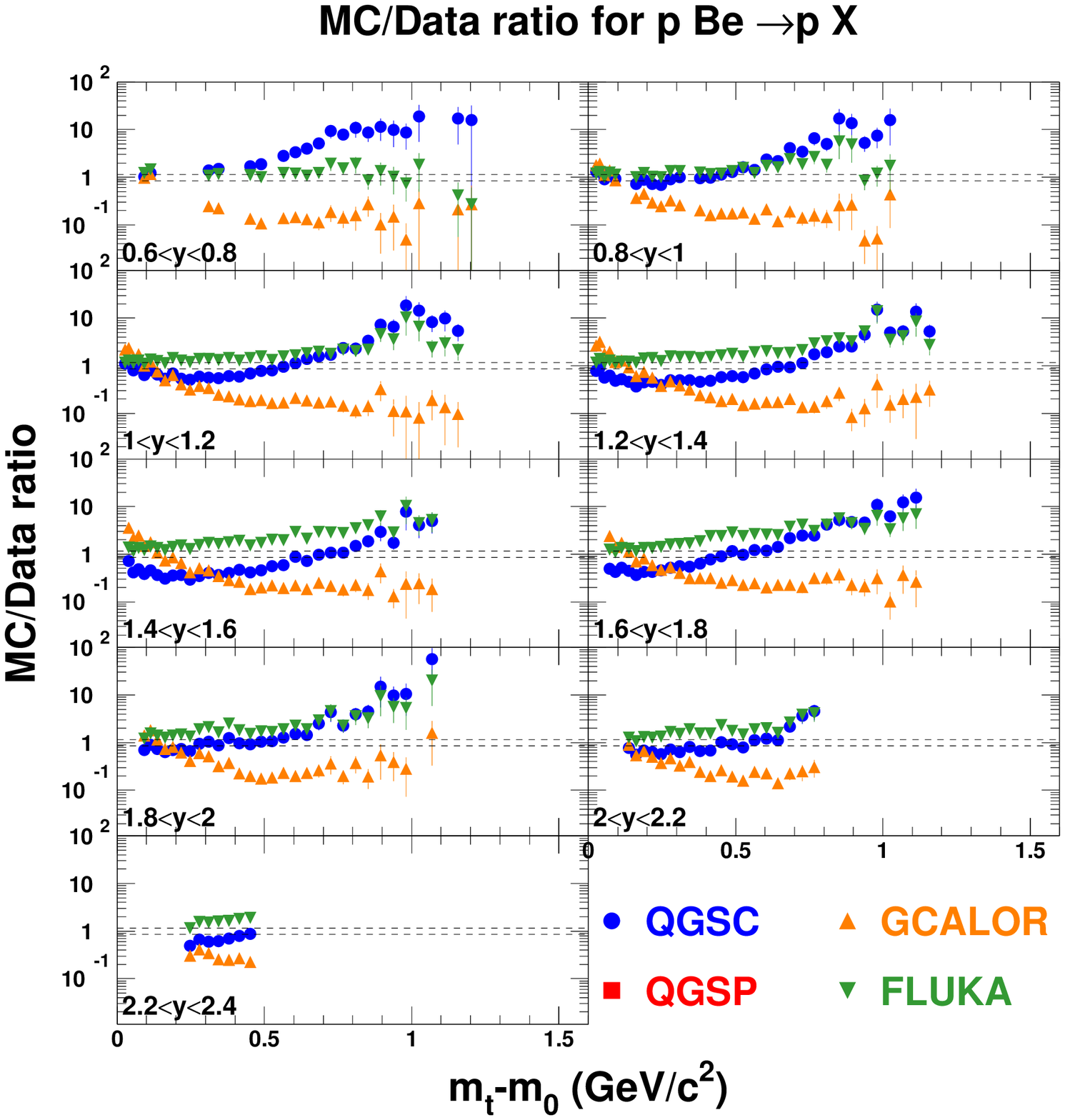,width=.495\linewidth}
\epsfig{file=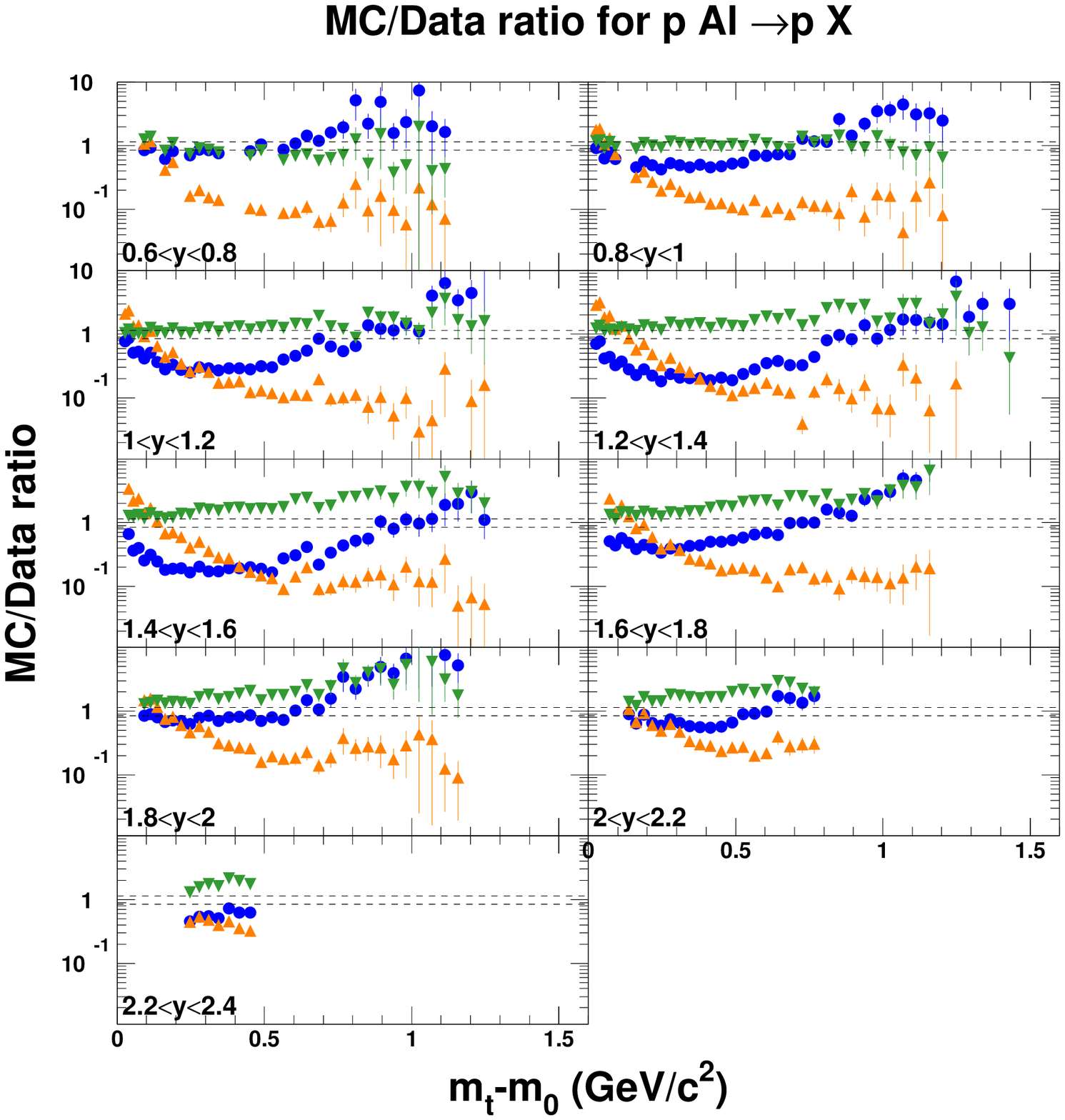,width=.495\linewidth}
\epsfig{file=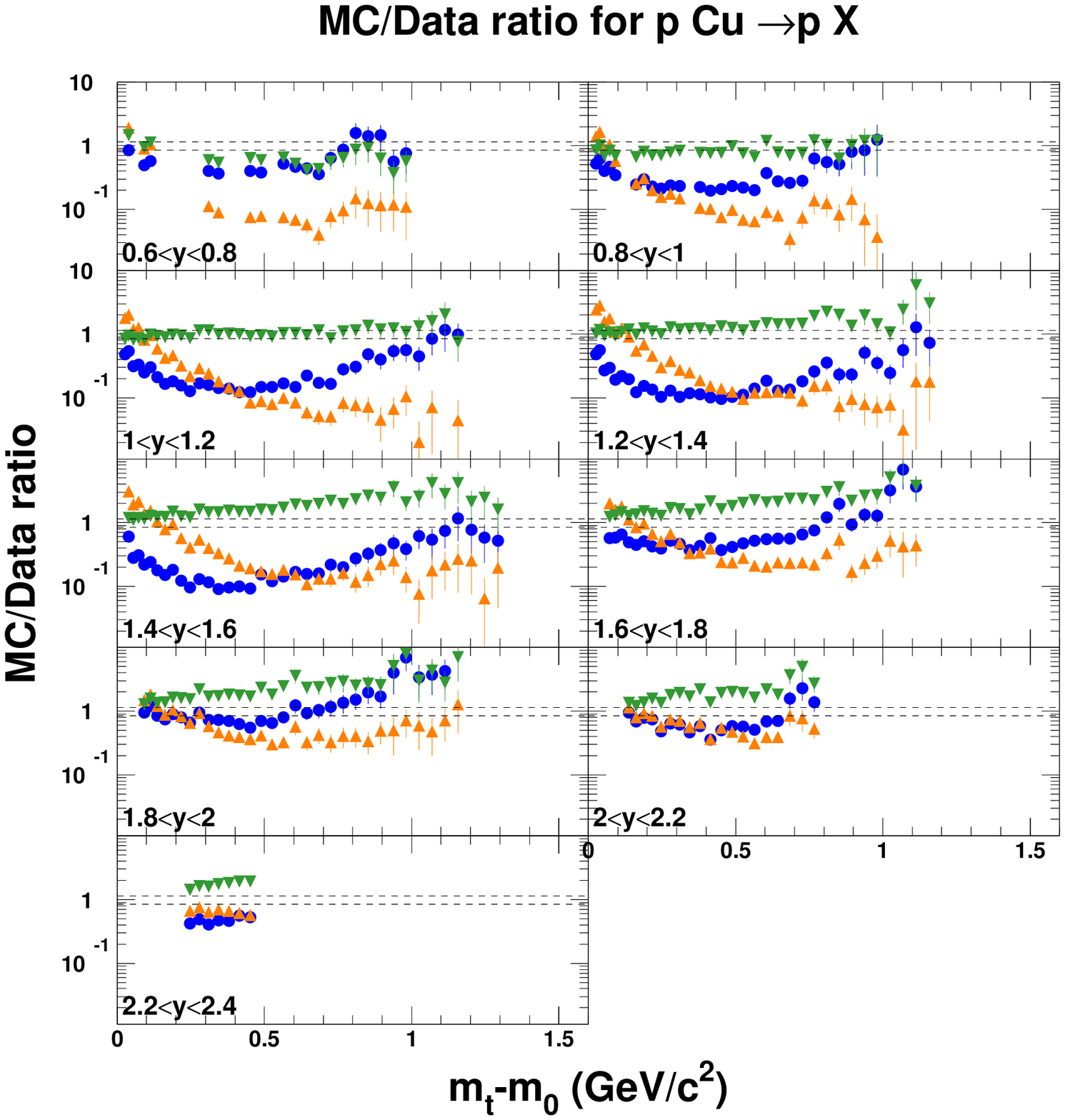,width=.495\linewidth}
\epsfig{file=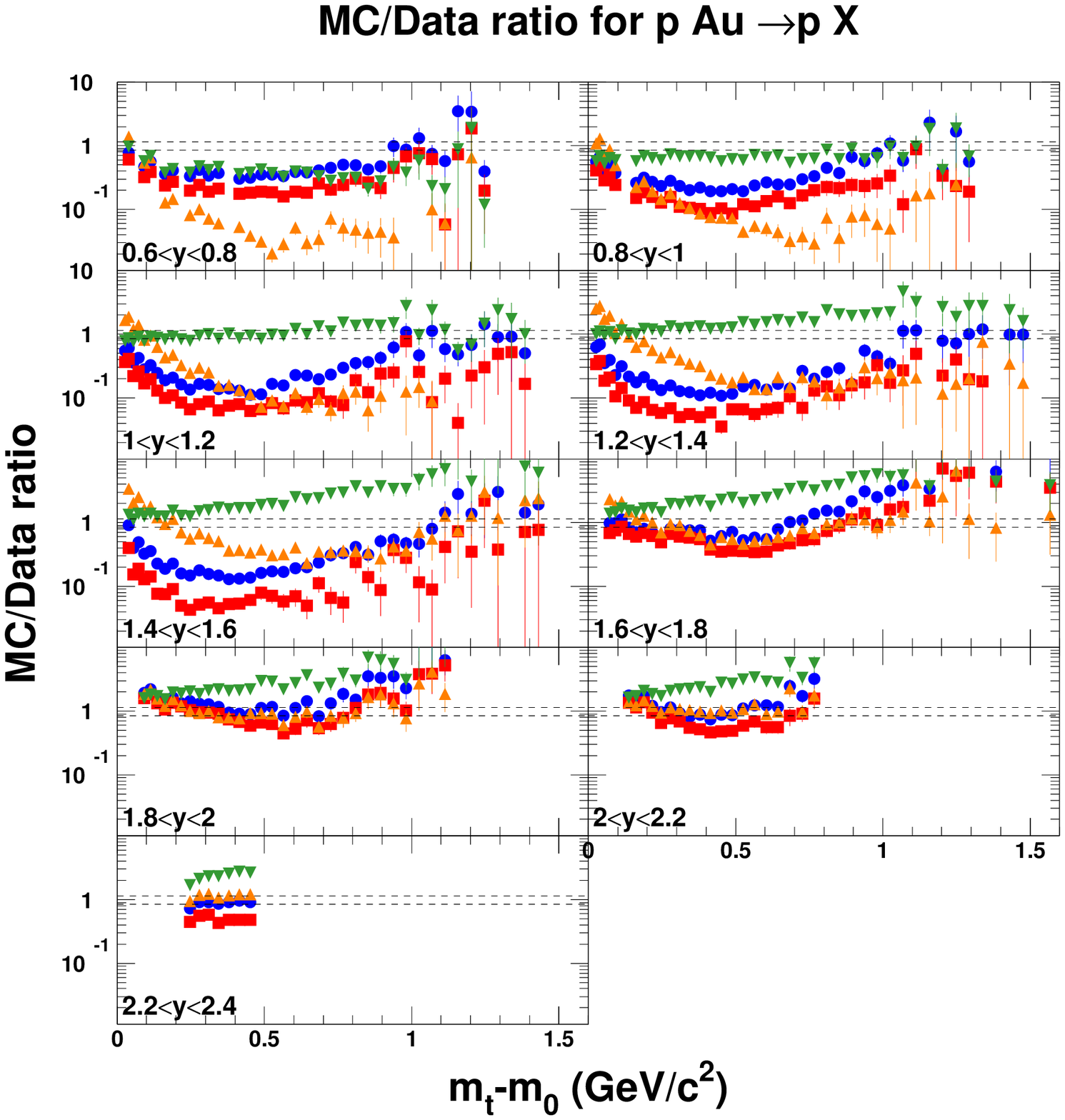,width=.495\linewidth}
\caption{\label{ratio.p}
The ratio of the simulated (MC) and data invariant cross sections  
as a function of transverse kinetic energy $m_{\rm t} - m_0$ 
in 0.2 bins of rapidity for the  
proton data
for $p$-Be, $p$-Al, $p$-Cu, and $p$-Au collisions.
The horizontal dashed lines indicate the $\pm 15\%$ normalization uncertainty of the E802 data. 
Only the QGSC ratios are shown as all the GEANT4 models given consistent predictions 
except for the gold target. 
Note that the vertical upper limit for the $p$-Be ratios are an order of magnitude
larger than the other targets.}
\end{center}\end{figure}

 The data and simulation results 
for the invariant cross-sections 
are shown in 
Figures~\ref{overlay.pi+},
 \ref{overlay.pi-}, 
 \ref{overlay.K+}, 
 \ref{overlay.K-}, 
 \ref{overlay.p}, and
 \ref{overlay.d} for the 
four targets for $\pi^+$, $\pi^-$, $K^+$,  $K^-$, $p$, and $d$ data, respectively.
Only the QGSC GEANT4 results are shown for the $\pi^\pm$, $K^\pm$, and $p$ data as 
the four GEANT4 simulation packages give nearly identical results. 
The statistical uncertainties in the results from simulation are similar for
all models and are only shown for the QGSC package to aid comparison with
the statistical uncertainties in the data. 
For the deuteron data, only the packages that give non-zero cross sections
for $y>0.4$ are shown in Figure~\ref{overlay.d}.
The ratios of Monte Carlo results to data are shown
in Figures~\ref{ratio.pi+},
 \ref{ratio.pi-}, 
 \ref{ratio.K+}, 
 \ref{ratio.K-}, and \ref{ratio.p} 
for 
all four targets for $\pi^+$, $\pi^-$, $K^+$,  $K^-$, and $p$, respectively.
The ratios are not shown for  the deuteron data
given the sparse nature and obviously poor agreement with the simulation.
There is no significant difference in the ratios for $\pi^\pm$, $K^\pm$ and $p$ production
for the four GEANT4 models except for QGSC and QGSP for the heaviest target; 
accordingly we show only the QGSC prediction except for the gold target.
The predictions for the QGSP,  QGSP\_BIC, and QGSC\_LEAD\_HP models are
statistically consistent for all targets.
The statistical uncertainty in the data and the Monte Carlo are combined
in quadrature to produce the uncertainty in the ratios shown in the Figures
while the E802 normalization uncertainty is indicated separately.

As seen in Figure~\ref{ratio.pi+}
for $\pi^+$ production, FLUKA generally has good agreement in slope but
overestimates the magnitude by up to a factor of two at low rapidity.
All the GEANT4 packages give similar results and agree in magnitude with
the data at lowest $m_{\rm t}$ but do not agree in slope for $y<$ 1.4, 1.6, 1.6, and 1.8
for the Be, Al, Cu, and Au targets, respectively. GCALOR has better agreement
than FLUKA or GEANT4 for the  $\pi^+$ data.

Similar observations can be made for $\pi^-$ production in Figure~\ref{ratio.pi-}.
GCALOR most accurately reproduces the data over the measured kinematic range
with some underestimate of the magnitude at low  $m_{\rm t}$ and high rapidity.
FLUKA has reasonable agreement in slope and is within a 
factor of two in magnitude for all the data.
The GEANT4 agreement is good for all the Be data and
agrees for the heavier targets at low $m_{\rm t}$ or $y>$ 1.2, 1.6, and 1.6 for
Al, Cu, and Au, respectively.

For positive kaon production (Figure~\ref{ratio.K+}), FLUKA agrees in slope
and magnitude for the Al, Cu, and Au targets. For the Be target, FLUKA agrees
in slope but the magnitude is higher than the data. 
The GCALOR agreement with the data is comparable to FLUKA for the Be target, but
consistently underestimates the magnitude for the heavier targets.
All the GEANT4
packages have the wrong slope for all targets and only agree in magnitude
at  lowest $m_{\rm t}$. 

Both FLUKA and GCALOR reproduce the slope of the  $K^-$ data reasonably well (Figure~\ref{ratio.K-}).
The magnitude predicted by FLUKA is higher than the data, while GCALOR has better agreement.
The slope of the Be data is reproduced reasonably well by the GEANT4 packages,
but is lower in magnitude than the data. For the heavier targets, GEANT4
predicts a slope less than that of the data and agrees in magnitude only at 
lowest  $m_{\rm t}$. 

The ratios of the Monte Carlo results to the data 
 for proton production are shown in Figure~\ref{ratio.p}.
For the Be target, both FLUKA and the GEANT4 packages have a slope greater than that of the
data  with moderately good agreement with the data in magnitude at low  $m_{\rm t}$.
For the heavier targets, FLUKA generally has good agreement for $y < 1.3$, but the
predicted slope exceeds the data for larger rapidities.
For the GEANT4 packages for the heavier targets, the agreement is poor
for $y<$ 2.2, 1.6, and 1.6 for Al, Cu, and Au, respectively, but improves
somewhat at higher rapidities. In general the slope of the GEANT4 packages
does not match the data well over the full range of measured  $m_{\rm t}$.
It is notable that the greatest difference between QGSC and the other GEANT4 packages is for proton
production and for the
Au target. GCALOR has the poorest agreement with the data.

None of the simulation packages reproduces the deuteron data (Figure~\ref{overlay.d}) well.
Neither GCALOR nor FLUKA predict a significant production of deuterons for rapidity above 0.5.
QGSC and QGSC\_LEAD\_HP underestimate the deuteron rate by an order of magnitude but 
do a reasonable job at predicting the slope of the deuteron data at low rapidity.
The agreement is worse at rapidity greater than $\sim\!0.8$.


\section{Conclusions and discussion}
\label{sec:conclusions}

 The FLUKA simulation package gives the best overall agreement with
the E802 meson data with the greatest deviation between the data and Monte Carlo 
of a factor of $\sim\!2$ over the entire kinematic range of the data.
The agreement of the GEANT3 and GEANT4 packages was worse in general.
The agreement of all the simulation packages with the proton and deuteron
production data was less satisfactory than that for the meson data.

 We note that a previous investigation with  the JAM~\cite{ref:JAM} 
simulation package gave good agreement and that JAM has been interfaced with
GEANT4~\cite{Koi:2003iq} although there is no current plan to implement JAM as a hadronic
physics list in GEANT4. In addition a great deal of data with similar targets
and kinematics is currently being analyzed or accumulated~\cite{Barr:2005gd} and
should provide for validation and refinement of simulations.

\section{Acknowledgements}
\label{sec:acknowledgements}

 We wish to thank Andrei Poblaguev for useful conversations and suggestions
based on his earlier, unpublished comparisons of simulations with E802 data.
We also thank Peter Gumplinger for assistance with the GEANT4 simulation.
We acknowledge the assistance of E802 collaborators Dana Beavis, Chellis Chasman
and Ramiro Debbe 
and Boris Pritychenko of the National Nuclear Data Center 
in locating the tables of the E802 results.

This manuscript was authored by Brookhaven Science Associates, LLC under 
Contract No. DE-AC02-98CH1-886 with the U.S. Department of Energy.
The United States Government retains, and the publisher, by 
accepting the article for publication,
acknowledges, a world-wide license to publish or reproduce 
the published form of this manuscript, or
allow others to do so, for the United States Government purposes.

This work was partially funded by  National Science Foundation Grant 
\#0428662 to NYU for RSVP Advanced Planning.



\clearpage 

\end{document}